\documentclass[sigconf]{acmart}

\usepackage{amssymb}
\usepackage{amsmath}
\usepackage{booktabs}
\usepackage{color}
\usepackage{listings}
\usepackage{mathtools}
\usepackage{pdfpages}
\usepackage{soul} % Highlighting text
\usepackage{subcaption}

\usepackage{pifont}
\newcommand{\yescheck}{\ding{51}}
\newcommand{\nocheck}{\ding{55}}

\definecolor{darkgreen}{RGB}{0,127,0}
\definecolor{darkred}{RGB}{127,0,0}
\definecolor{shadecolor}{RGB}{250, 250, 250}
\definecolor{lightyellow}{RGB}{250, 250, 212}
\definecolor{HLYELLOW}{RGB}{240, 127, 0}
\definecolor{hlyellow}{RGB}{240, 127, 0}
\sethlcolor{white}

\newcommand{\yin}{\includegraphics[height=0.8em]{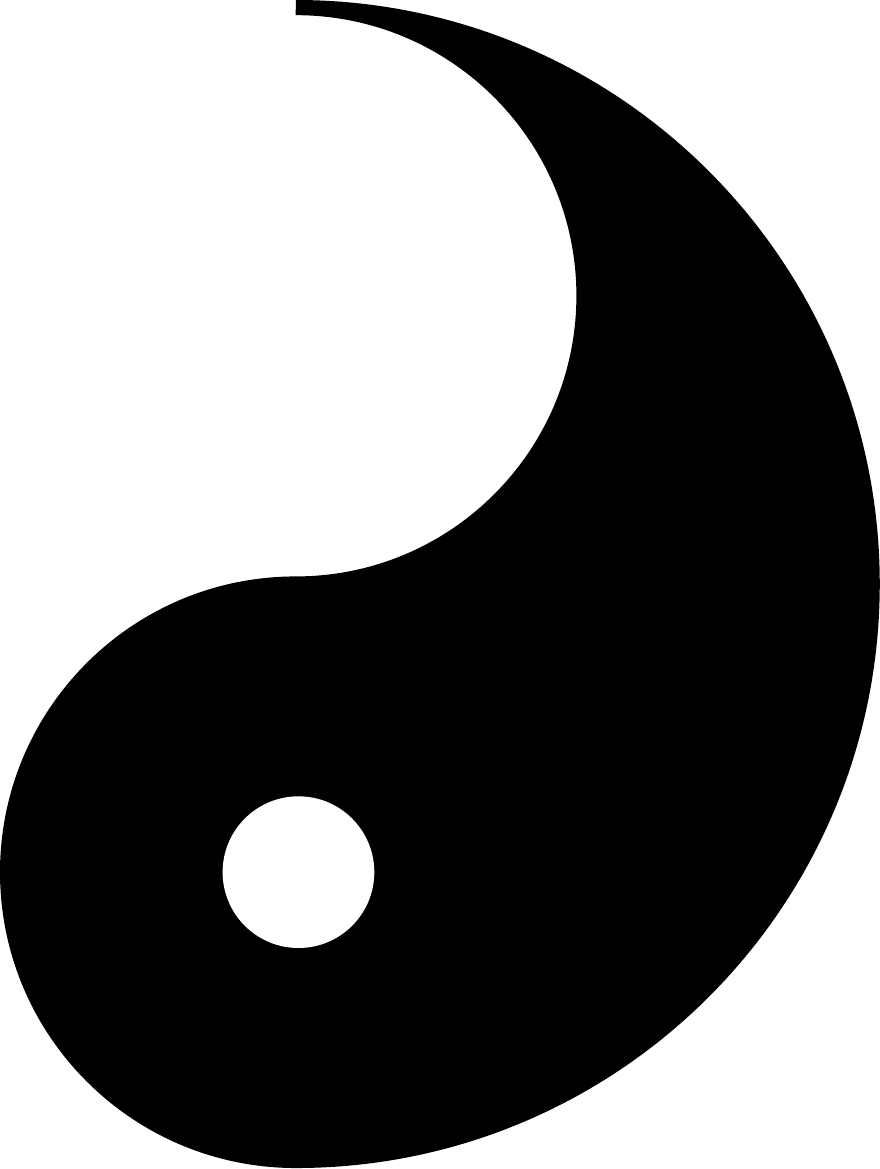}}
\newcommand{\yang}{\includegraphics[height=0.8em]{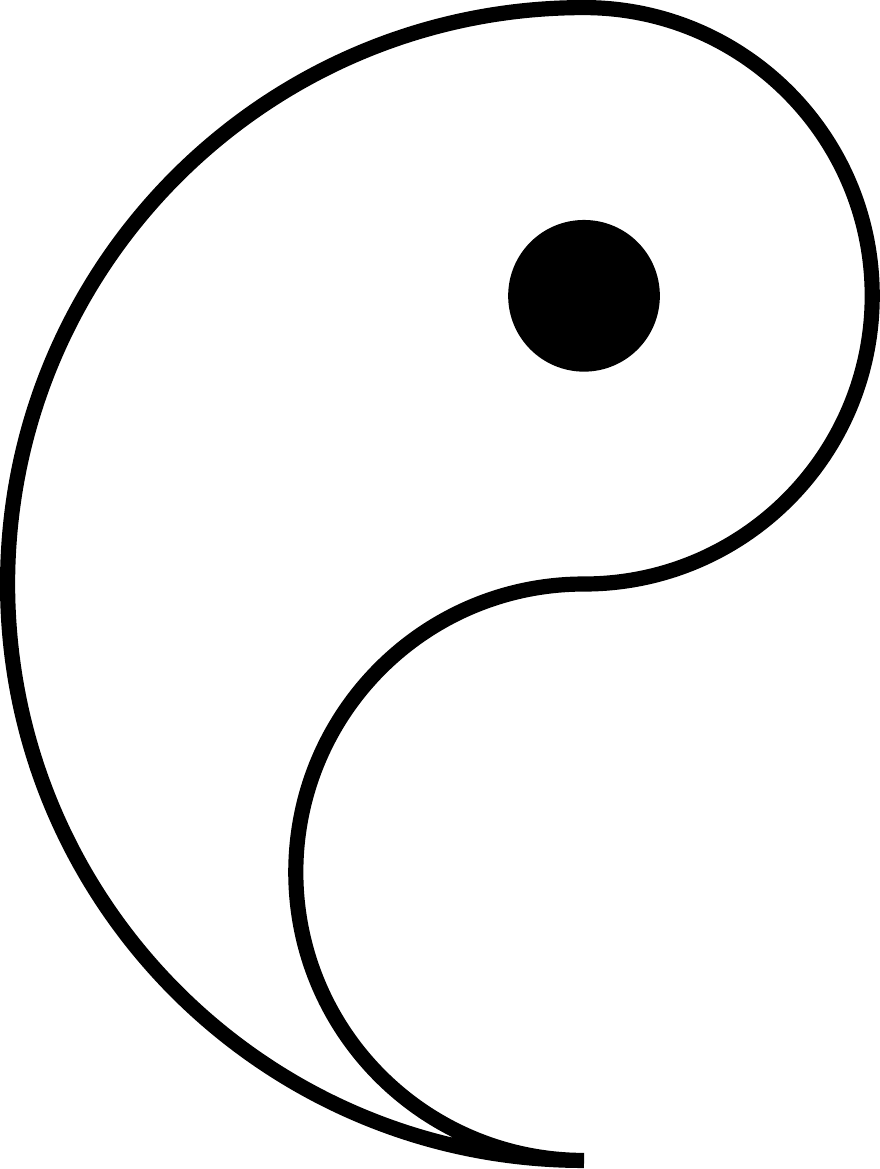}}
\newcommand{\yinyang}{\includegraphics[height=0.8em]{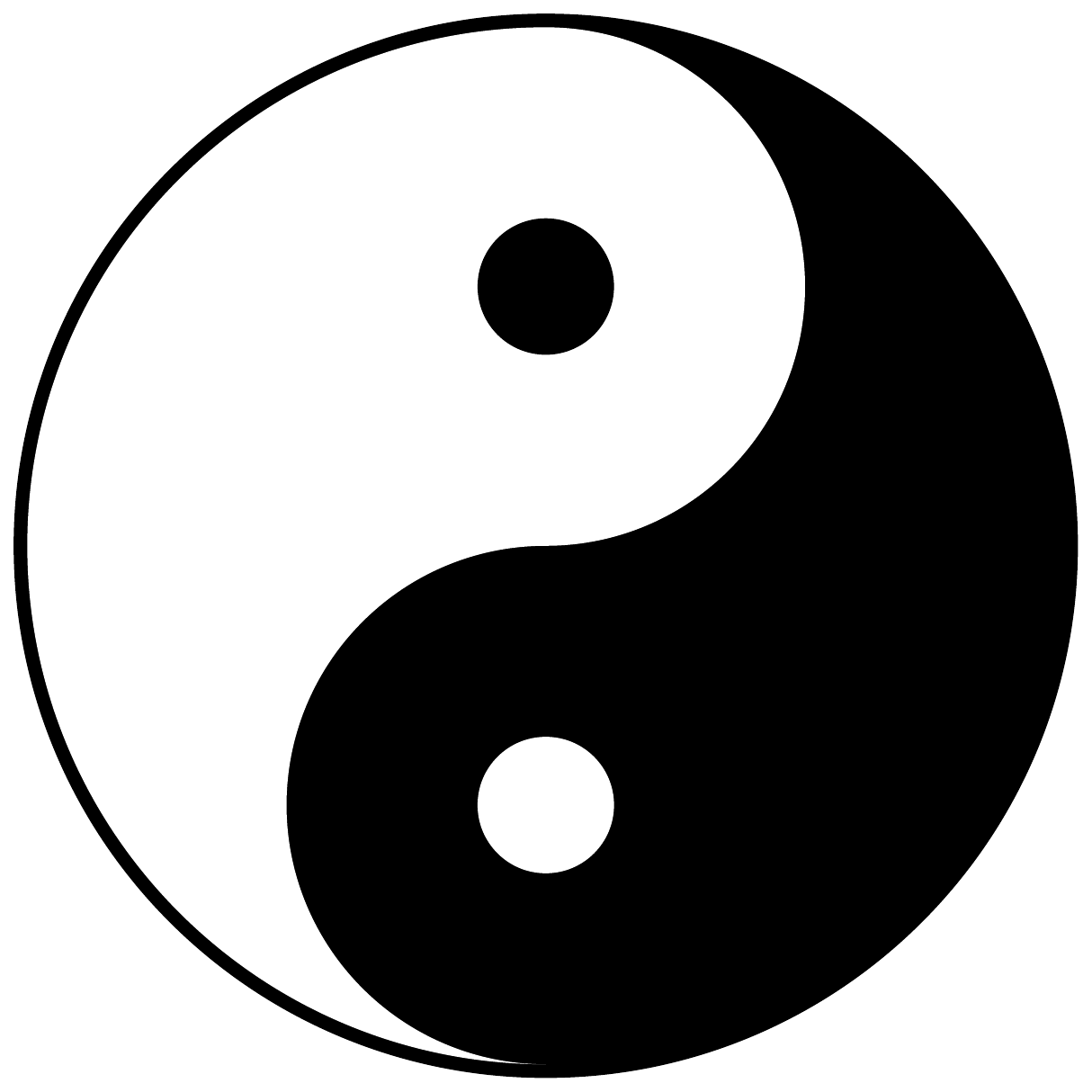}}

\def\BibTeX{{\rm B\kern-.05em{\sc i\kern-.025em b}\kern-.08emT\kern-.1667em\lower.7ex\hbox{E}\kern-.125emX}}
    
\copyrightyear{2019} 
\acmYear{2019} 
\acmConference[SC '19]{The International Conference for High Performance Computing, Networking, Storage, and Analysis}{November 17--22, 2019}{Denver, CO, USA}
\acmBooktitle{The International Conference for High Performance Computing, Networking, Storage, and Analysis (SC '19), November 17--22, 2019, Denver, CO, USA}
\acmPrice{15.00}
\acmDOI{10.1145/3295500.3356200}
\acmISBN{978-1-4503-6229-0/19/11}

\hypersetup{final}

\setcopyright{none}

\begin{document}

\newcommand{\todo}[1]{{\color{red} #1}}

\title[Optimizing Quantum Transport Simulations via Data-Centric Parallel Programming]{Optimizing the Data Movement in Quantum Transport Simulations via Data-Centric Parallel Programming}

\author{Alexandros Nikolaos Ziogas$^{*}$, Tal Ben-Nun$^{*}$, Guillermo Indalecio Fern\'{a}ndez$^{\scriptscriptstyle\dagger}$,
  Timo Schneider$^{*}$, Mathieu Luisier$^{\scriptscriptstyle\dagger}$, and Torsten Hoefler$^{*}$} 
\affiliation{%
$^*$Scalable Parallel Computing Laboratory, ETH Zurich, Switzerland\\
$^{\dagger}$Integrated Systems Laboratory, ETH Zurich, Switzerland 
}

\renewcommand{\shortauthors}{Ziogas et al.}

\begin{abstract}
Designing efficient cooling systems for integrated circuits (ICs)
relies on a deep understanding of the electro-thermal properties of
transistors. To shed light on this issue in currently fabricated
FinFETs, a quantum mechanical solver capable of revealing
atomically-resolved electron and phonon transport phenomena from
first-principles is required. In this paper, we consider a global,
data-centric view of a state-of-the-art quantum transport simulator to
optimize its execution on supercomputers. The approach yields coarse-
and fine-grained data-movement characteristics, which are used for
performance and communication modeling, communication-avoidance,
and data-layout transformations. The transformations are
tuned for the Piz Daint and Summit supercomputers, where 
each platform requires different caching and fusion
strategies to perform optimally. The presented results make \textit{ab 
initio} device simulation enter a new era, where nanostructures
composed of over 10,000 atoms can be investigated at an unprecedented
level of accuracy, paving the way for better heat management in
next-generation ICs.
\end{abstract}



\maketitle

\begin{figure}
\centering
\includegraphics[width=\linewidth]{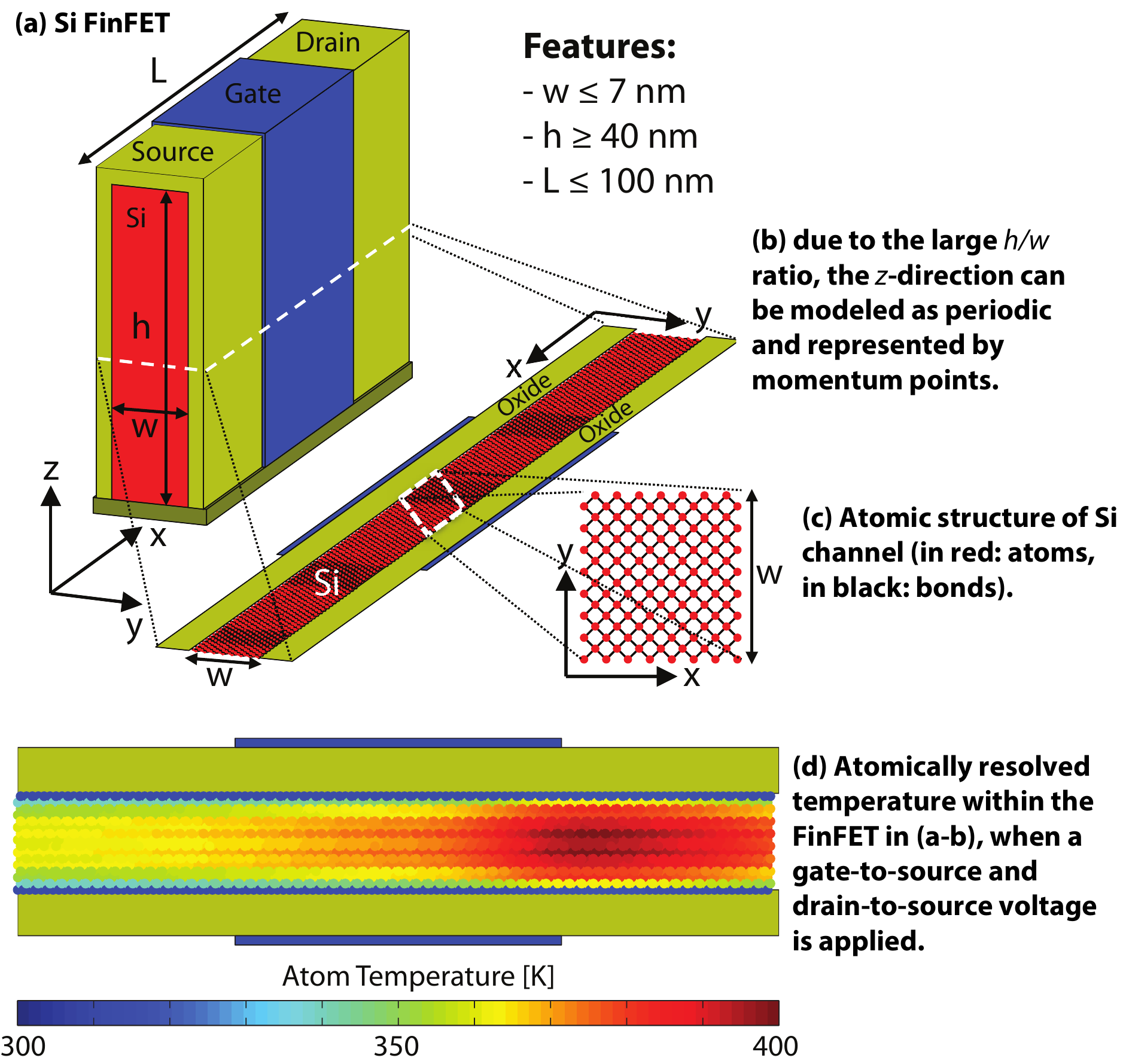} 
\caption{Schematic view of a 3-D Silicon FinFET.}
\label{fig:res}
\end{figure}

\section{Introduction}
Heat dissipation in microchips reached alarming peak values of 100
W/cm$^2$ already in
2006~\cite{wei,pop}. This led to the end of Dennard scaling and the
beginning of the ``multicore crisis'', an era with energy-efficient
parallel, but sequentially slower multicore CPUs.
Now, more than ten years later, average power densities of up to 30
W/cm$^2$, about four times more than hot plates, are commonplace in
modern high-performance CPUs, putting thermal management at the center
of attention of circuit designers~\cite{hotplate}. 
By scaling the dimensions of transistors more rapidly than their supply
voltage, the semiconductor industry has kept increasing heat dissipation
from one generation of microprocessors to the other. 
In this context, large-scale data and supercomputing centers are facing
critical challenges regarding the design and cost of their cooling
infrastructures. 
The price to pay for that has become exorbitant, as the cooling can be up
to 40\% of the total electicity consumed by data centers; a cumulative
cost of many billion dollars per year.

Landauer's theoretical limit of energy consumption for non-reversible
computing offers a glimmer of hope: today's processing units require
orders of magnitude more energy than the $k_BT\ln 2$ Joule bound to (irreversibly) change
one single bit. However, to approach this limit, it will be necessary to
first properly understand the mechanisms behind nanoscale heat
dissipation in semiconductor devices~\cite{pop}. Fin field-effect
transistors (FinFETs), as schematized in Fig.~\ref{fig:res}(a-c), build
the core of all recent integrated circuits (ICs). Their dimensions do not
exceed 100 nanometers along all directions, even 10 nm along one of
them, with an active region composed of fewer than 1 million
atoms. This makes them subject to strong quantum mechanical and
peculiar thermal effects.

When a voltage is applied across FinFETs, electrons start to flow from
the source to the drain contact, giving rise to an electrical
current whose magnitude depends on the gate bias. The potential
difference between source and drain allows electrons to transfer part
of their energy to the crystal lattice surrounding them. This energy
is converted into atomic vibrations, called phonons, that can propagate
throughout FinFETs. The more atoms vibrate, the ``hotter'' a device
becomes. This phenomenon, known as self- or Joule-heating, plays a
detrimental role in today's transistor technologies and has
consequences up to the system level. It is illustrated in
Fig.~\ref{fig:res}(d): a strong increase of the lattice temperature
can be observed close to the drain contact of the simulated
FinFET. The negative influence of self-heating on the CPU/GPU
performance can be minimized by devising computer-assisted strategies
to efficiently evacuate the generated heat from the active region of
transistors.

Electro-thermal properties of nano-devices can be modeled and
analyzed via \textit{Quantum Transport (QT) simulation}, where
electron and phonon currents are evaluated by taking quantum mechanics into account.
Due to the large height/width ratio of FinFETs, these effects can be 
physically captured in a two-dimensional simulation domain 
comprising 10,000 to 15,000 thousand atoms.
Such dissipative simulations
involve solving the Schr\"{o}dinger equation with open boundary
conditions over several momentum and energy vectors that are coupled
to each other through electron-phonon interactions.
A straightforward algorithm to address this numerical problem consists of
defining two loops, one over the momentum points and another one over the
electron energies, which results in potentially extraneous execution dependencies and
a complex communication pattern. The latter scales sub-optimally with
the number of participating atoms and computational resources, thus
limiting current simulations to the order of thousand atoms.

While the schedule (ordering) of the loops in the solver is natural 
from the physics perspective (\S~\ref{sec:problem}), the data decomposition it imposes 
when parallelizing is not scalable from the computational perspective.
To investigate larger device structures, it is crucial to
reformulate the problem as a communication-avoiding algorithm ---
rescheduling computations across compute resources to minimize data
movement. %--- creating a scalable communication scheme at the expense

Even when each node is operating at maximum efficiency, large-scale QT simulations are both bound by communication volume and
memory requirements.
The former inhibits strong scaling, as simulation time
includes nanostructure-dependent point-to-point communication patterns,
which becomes infeasible when increasing node count.
The memory bottleneck is a direct result of the former. It  hinders
large simulations due to the increased memory requirements w.r.t. atom count. 
Transforming the QT simulation algorithm to minimize communication is
thus the key to simultaneously model larger devices and increase
scalability on different supercomputers.

The current landscape of supercomputing resources is dominated by heterogeneous nodes, 
where no two clusters are the same. 
Each setup requires careful tuning of application performance, focused mostly 
around data locality~\cite{padal}. 
As this kind of tuning demands in-depth knowledge of the hardware, it is typically 
performed by a \yang~\textit{Performance Engineer}, a developer who is versed 
in intricate system details, existing high-performance libraries, 
and capable of modeling performance and setting up optimized
procedures independently.
This role, which complements the \yin~\textit{Domain Scientist}, has been 
increasingly important in scientific computing
for the past three decades, but is now essential for any application 
beyond straightforward linear algebra to operate at extreme scales. 
Until recently, both Domain Scientists and Performance Engineers would work 
with one code-base. This creates a co-dependent \yinyang~situation~\cite{pat-maccormicks-sos-talk},
where the original domain code is tuned to a point that making
modifications to the algorithm or transforming its behavior is
difficult to one without the presence of the other, even if data
locality or computational semantics are not changed. 

In this paper, we propose a paradigm change by rewriting the problem
from a data-centric perspective. 
We use OMEN, the current state-of-the-art quantum transport simulation
application~\cite{omen}, as our baseline, and show that the key to
formulating a communication-avoiding variant is tightly coupled with
recovering local and global data dependencies of the application.
We start from a reference Python implementation, using Data-Centric
(DaCe) Parallel Programming~\cite{sdfg} to express the computations
separately from data movement. DaCe automatically constructs a
\textit{stateful dataflow} view that can be used to optimize data
movement without modifying the original computation. This enables
rethinking the communication pattern of the simulation, and tuning the
data movement for each target supercomputer. 

In sum, the paper makes the following contributions:
\begin{itemize}
	\item[\yin]
          Construction of the dissipative quantum transport simulation problem from a physical perspective; % domain
	\item[\yin]
          Definition of the stateful dataflow of the algorithm, making
          data and control dependencies explicit on all levels;
	\item[\yang]
          Creation of a novel \textit{tensor-free
            communication-avoiding variant} of the algorithm based on the data-centric view; 
	\item[\yang]
          Optimal choice of the decomposition parameters based on the
          modeling of the performance and communication of our
          variant, nano-device configuration, and cluster architecture;
	\item[\yinyang]
          Demonstration of the algorithm's scalability
          on two vastly different supercomputers --- Piz Daint
          and Summit --- up to full-scale runs on 10k atoms, 21 momentum points and 1,000 energies per momentum;
	\item[\yinyang]
          A performance increase of \textit{1--2 orders of magnitude} over the previous
          state of the art, all from a data-centric Python implementation that
          reduces code-length by a factor of five.
\end{itemize}

\section{Statement of the Problem}
\label{sec:problem}

A technology computer aided design (TCAD) tool can shed light on the
electro-thermal properties of nano-devices, provided that it includes
the proper physical models:
\begin{itemize}
	\item The dimensions of FinFETs calls for an atomistic quantum 
	mechanical treatment of the device structures;
	\item The electron and phonon bandstructures should
	be accurately and fully described;
	\item The interactions between electrons and phonons, especially
	energy exchanges, should be accounted for.
\end{itemize}
The Non-equilibrium Green's Function (NEGF) formalism \cite{datta}
combined with density functional theory (DFT) \cite{kohn} fulfills all
these requirements and lends itself optimally to the investigation of
self-heating in arbitrary device geometries. With NEGF, both electron
and phonon transport can be described, together with their respective
interactions. With the help of DFT, an \textit{ab initio} method, any
material (combination) can be handled at the atomic level without the
need for empirical parameters.  

FinFETs are essentially three-dimensional (3-D) components that can be
approximated as 2-D slices in the $x$-$y$ plane, whereas the height,
aligned with the $z$-axis (see Fig.~\ref{fig:res}(a-b)), can be
treated as a periodic dimension and represented by a momentum vector
$k_z$ or $q_z$ in the range $[-\pi, \pi]$. Hence, the DFT+NEGF
equations have the following form:
\begin{eqnarray}
\left\{
\begin{array}{l}
\left(E\cdot{\mathbf S}(k_z)-{\mathbf H}(k_z)-{\mathbf \Sigma}^{R}(E,k_z)\right)\cdot 
{\mathbf G}^{R}(E,k_z)={\mathbf I}\\
{\mathbf G}^{\gtrless}(E,k_z)=\mathbf{G}^{R}(E,k_z)\cdot{\mathbf
  \Sigma}^{\gtrless}(E,k_z)\cdot{\mathbf G}^{A}(E,k_z).
\end{array}
\right.
\label{eq:1}
\end{eqnarray}
In Eq.~(\ref{eq:1}), $E$ is the electron energy, ${\mathbf S}(k_z)$
and ${\mathbf H}(k_z)$ the overlap and Hamiltonian matrices,
respectively. They typically exhibit a block tri-diagonal structure
and a size $N_A\times N_{orb}$ with $N_A$ as the total number of atoms
in the considered structure and $N_{orb}$ the number of orbitals
(basis components) representing each atom. The ${\mathbf S}(k_z)$
and ${\mathbf H}(k_z)$ matrices must be produced by a DFT package
relying on a localized basis set, e.g., SIESTA \cite{siesta} or CP2K
\cite{cp2k}. The ${\mathbf G}(E,k_z)$'s refer to the electron Green's
Functions (GF) at energy $E$ and momentum $k_z$. They are of the same
size as ${\mathbf S}(k_z)$, ${\mathbf H}(k_z)$, and ${\mathbf I}$, the
identity matrix. The GF can be either retarded ($R$), advanced ($A$), lesser
($<$), or greater ($>$) with ${\mathbf G}^A(E,k_z)$=$\left({\mathbf
  G}^R(E,k_z)\right)^T$. The same conventions apply to the 
self-energies ${\mathbf\Sigma}(E,k_z)$ that include a boundary and a
scattering term. The former connects the simulation domain to
external contacts, whereas the latter encompasses all possible
interactions of electrons with their environment. 

To handle phonon transport, the following NEGF-based system of
equations must be processed:
\begin{eqnarray}
\left\{
\begin{array}{l}
\left(\omega^2\cdot{\mathbf I}-{\mathbf \Phi}(q_z)-{\mathbf \Pi}^{R}(\omega,q_z)\right)
\cdot {\mathbf D}^{R}(\omega,q_z)={\mathbf I}\\
{\mathbf D}^{\gtrless}(\omega,q_z)={\mathbf D}^{R}(\omega,q_z)
\cdot {\mathbf \Pi}^{\gtrless}(\omega,q_z)\cdot{\mathbf D}^{A}(\omega,q_z),
\end{array}
\right.
\label{eq:2}
\end{eqnarray}
where the ${\mathbf D}(\omega,q_z)$'s are the phonon Green's functions
at frequency $\omega$ and momentum $q_z$ and the ${\mathbf
  \Pi}(\omega,q_z)$'s the self-energies, while ${\mathbf \Phi}(q_z)$
refers to the dynamical (Hessian) matrix of the studied domain,
computed with density functional perturbation theory (DFPT)
\cite{dfpt}. The phonon Green's Function types are the same as for
electrons (retarded, advanced, lesser, and greater). All
matrices involved in Eq.~(\ref{eq:2}) are of size $N_A\times N_{3D}$,
$N_{3D}$=3 corresponding to the number of directions along which the
crystal can vibrate ($x$, $y$, $z$).

\begin{figure}
	\centering
	\includegraphics[width=.8\linewidth]{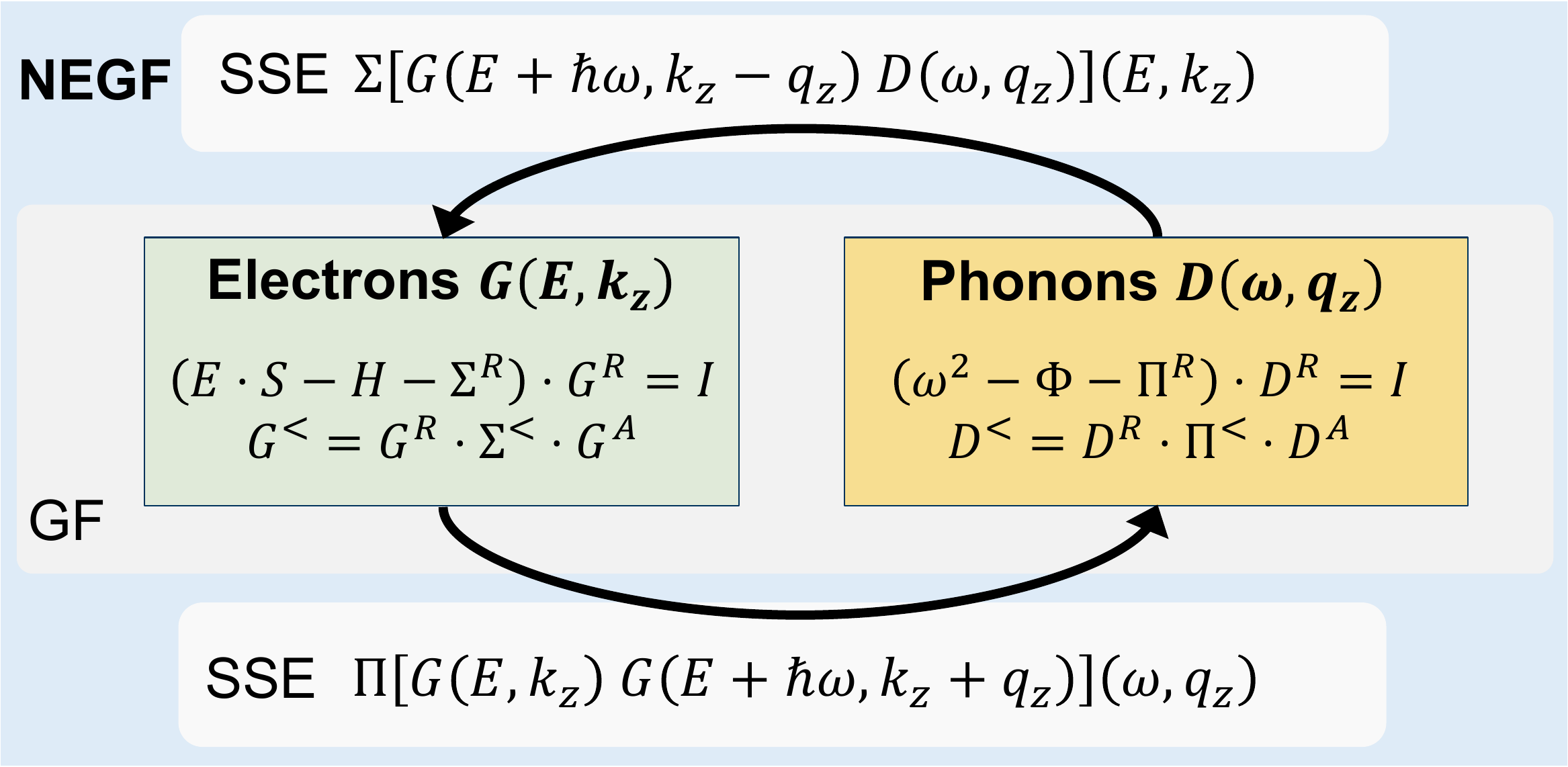} 
	\vspace{-1em}
	\caption{\hl{Self-consistent coupling between the GF and SSE phases (kernels) as
		part of the NEGF formalism.}}
	\vspace{-1em}
	\label{fig:gf_sse}
\end{figure}

Equations (\ref{eq:1}) and (\ref{eq:2}) must be solved for all
possible electron energy ($N_E$) and momentum ($N_{k_z}$) points as
well as all phonon frequencies ($N_{\omega}$) and momentum
($N_{q_z}$). This can be done with a so-called recursive Green's
Function (RGF) algorithm \cite{rgf} that takes advantage of the block
tri-diagonal structure of the matrices ${\mathbf H}$, ${\mathbf S}$,
and ${\mathbf \Phi}$. All matrices can be divided into $bnum$ blocks with
$\frac{N_A}{bnum}$ atoms each, if the structure is homogeneous, as
here. RGF then performs a forward and backward pass over the $bnum$
blocks that compose the 2-D slice. Both passes involve a number of
multiplications between matrices of size
$\left(\frac{N_A}{bnum}N_{orb}\right)^2$ for electrons (or
$\left(\frac{N_A}{bnum}N_{3D}\right)^2$ for phonons) for each block.

The main computational bottleneck does not come from RGF, but from the fact that in
the case of self-heating simulations the energy-momentum ($E$,$k_z$)
and frequency-momentum ($\omega$,$q_z$) pairs are not independent from
each other, but tightly coupled through the scattering self-energies
(SSE) ${\mathbf \Sigma}^{R\gtrless,S}$ and ${\mathbf
  \Pi}^{R\gtrless,S}$. These matrices are made of blocks of size
$N_{orb}\times N_{orb}$ and $N_{3D}\times N_{3D}$, respectively,
and are given by \cite{stieger}:
{\small
\begin{eqnarray}
{\mathbf \Sigma}^{\gtrless S}_{aa}(E,k_z)=i\sum_{q_zijl}\int
\frac{d\hbar\omega}{2\pi}\left[\nabla_i{\mathbf H}_{ab}\cdot 
{\mathbf G}^{\gtrless}_{bb}(E-\hbar\omega,k_z-q_z)\cdot\right.\nonumber\\
\nabla_j{\mathbf H}_{ba}\cdot\left({\mathbf D}^{\gtrless ij}_{ba}(\omega,q_z)-
{\mathbf D}^{\gtrless ij}_{bb}(\omega,q_z)-\right.\nonumber\\
\left.\left.{\mathbf D}^{\gtrless ij}_{aa}(\omega,q_z)+{\mathbf D}^{\gtrless
  ij}_{ab}(\omega,q_z)\right)\right],
\label{eq:3}
\end{eqnarray}
\begin{eqnarray}
{\mathbf \Pi}^{\gtrless S}_{aa}(\omega,q_z)=-i\sum_{k_zl}\int\frac{dE}{2\pi} 
\mathrm{tr}\left\{\nabla_{i}{\mathbf H}_{ba}\cdot{\mathbf G}^{\gtrless}_{aa}(E+\hbar\omega,k_z+q_z)\cdot\right.\nonumber\\
&\hspace*{-6cm}\left.\nabla_j{\mathbf H}_{ab}\cdot{\mathbf G}^{\lessgtr}_{bb}(E,k_z)\right\},
\label{eq:4}
\end{eqnarray}
\begin{eqnarray}
{\mathbf\Pi}^{\gtrless S}_{ab}(\omega,q_z)=i\sum_{k_z}\int\frac{dE}{2\pi} 
\mathrm{tr}\left\{\nabla_{i}\mathbf{H}_{ba}\cdot\mathbf{G}^{\gtrless}_{aa}(E+\hbar\omega,k_z+q_z)\cdot\right.\nonumber\\
&\hspace*{-6cm}\left.\nabla_j\mathbf{H}_{ab}\cdot\mathbf{G}^{\lessgtr}_{bb}(E,k_z)\right\}.
\label{eq:5}
\end{eqnarray}
}
In Eqs.~(\ref{eq:3}-\ref{eq:5}), all Green's Functions ${\mathbf
  G_{ab}}$ (${\mathbf D_{ab}}$) are matrices of size $N_{orb}\times
N_{orb}$ ($N_{3D}\times N_{3D}$). They describe the coupling between
  all orbitals (vibrational directions) of two neighbor atoms $a$ and
  $b$ situated at position ${\mathbf R}_a$ and 
  ${\mathbf R}_b$. Each atom possesses $N_B$ neighbors. Furthermore,
$\nabla_i {\mathbf H}_{ab}$ is the derivative of the Hamiltonian
  block $\mathbf{H}_{ab}$ coupling atoms $a$ and $b$ w.r.t variations
  along the $i$=$x$, $y$, or $z$ coordinate of the bond
  $\mathbf{R}_{b}-\mathbf{R}_{a}$. To obtain the retarded components
  of the scattering self-energies, the following relationship can be
  used: ${\mathbf \Sigma}^R\approx({\mathbf \Sigma}^{>}-{\mathbf
    \Sigma}^{<})/2$, which is also valid for ${\mathbf
    \Pi}^R$\cite{lake}. Due to computational reasons, only the
  diagonal blocks of ${\mathbf \Sigma}^{R\gtrless,S}$ are retained,
  while $N_B$ non-diagonal connections are kept for ${\mathbf
    \Pi}^{R\gtrless,S}$.

\begin{table}
	%\begin{minipage}[t]{.39\linewidth}
		\centering
		\caption{Typical QT Simulation Parameters}
		\vspace{-1em}
		\footnotesize
		\begin{tabular}{l p{2.75cm} r} 
			\toprule
			\textbf{Variable} & \textbf{Description} & \textbf{Range} \\
			\midrule
			$N_{k_z}$ & Number of electron momentum points	& $[1, 21]$ \\
			$N_{q_z}$ & Number of phonon momentum points	& $[1, 21]$ \\			
			$N_E$ &  Number of energy points	& [700, 1500] \\
			$N_\omega$ & Number of phonon frequencies	& [10, 100] \\
			%$a$ & Atoms in a 2-D slice	&  [1, 1000] \\
			%This might give the false impression that this can be done, while
			%this is an objective
			$N_A$ & Total number of atoms per device structure & See Table ~\ref{tab:competitors}\\
			$N_B$ & Neighbors considered for each atom	& [4, 50] \\
			$N_{orb}$ & Number of orbitals per atom & [1, 30] \\
			$N_{3D}$ & Degrees of freedom for crystal vibrations & 3 \\
			\bottomrule
		\end{tabular}
		\label{tab:common-values}
	%\end{minipage}\hfill
\end{table}

\begin{table*}
	\caption{State of the Art Quantum Transport Simulators}
	\small
	\vspace{-1em}
	\begin{tabular}{l lll lll rc} 
		\toprule
		\textbf{Name} &	\multicolumn{6}{c}{\bf Maximum \# of Computed Atoms} & \multicolumn{2}{c}{\bf Scalability} \\	\cmidrule(l{2pt}r{2pt}){2-7}%\cmidrule(l{2pt}r{2pt}){8-9}\addlinespace
		&\multicolumn{3}{c}{\bf \hl{Tight-binding-like$^*$}} & \multicolumn{3}{c}{\bf \hl{DFT}} & Max. Cores & Using\\\cmidrule(l{2pt}r{2pt}){2-4}\cmidrule(l{2pt}r{2pt}){5-7} %& Max. DFT+SSE\\
		& \hl{$GF^\dagger_e$} & $GF^\dagger_{ph}$ & $GF+SSE$ & \hl{$GF^\dagger_e$} & $GF^\dagger_{ph}$ & \hl{$GF+SSE$} & (Magnitude) & GPUs \\%& Performance \\
		\midrule
		GOLLUM~\cite{ferrer2014gollum}&1k&1k&---&100&100&---&N/A&\nocheck\\
		Kwant~\cite{groth2014kwant}&10k&---&---&---&---&---&N/A&\nocheck\\
		NanoTCAD ViDES~\cite{vides}&10k&---&---&---&---&---&N/A&\nocheck\\
		QuantumATK~\cite{atk}&10k&10k&---&1k&1k&---&1k&\nocheck\\		
		TB\_sim~\cite{niquet}&100k&---&10k$^\ddagger$&1k&---&---&10k&\yescheck\\		
		NEMO5~\cite{nemo5}&100k&100k&10k$^\ddagger$&---&---&---&100k&\yescheck \\
		OMEN~\cite{omen}&\hl{100k (1.44 Pflop/s} \cite{sc11}) &100k&10k&\hl{10k}&\hl{10k (15 Pflop/s} \cite{sc15})&\hl{1k (0.16 Pflop/s)}&100k&\yescheck\\	
		\midrule
		\textbf{This work} & N/A & N/A & N/A & 10k & 10k & \textbf{\hl{10k (19.71 Pflop/s)}} & \textbf{1M} & \yescheck \\ %57.14
		\bottomrule
	\end{tabular}\\
	{\footnotesize $^*$: including Maximally-Localized Wannier Functions (MLWF), $\dagger$: Ballistic, $\ddagger$: Simplified.}\vspace{-1em}
	\label{tab:competitors}
\end{table*}

The evaluation of Eqs.~(\ref{eq:3}-\ref{eq:5}) does not require the
knowledge of all entries of the ${\mathbf G}$ and ${\mathbf D}$
matrices, but of two (lesser and greater) 5-D tensors of shape
$[N_{k_z}, N_E, N_A, N_{orb}, N_{orb}]$ for electrons and two 6-D
tensors of shape $[N_{q_z}, N_{\omega}, N_A, N_B + 1, N_{3D},
  N_{3D}]$ for phonons. Each $[k_z, E, N_A, N_{orb}, N_{orb}]$ and
$[q_z, \omega, N_A, N_B + 1, N_{3D},$ $N_{3D}]$ combination is
produced independently from the other by solving Eq.~(\ref{eq:1}) and
(\ref{eq:2}), respectively. The electron and phonon scattering
self-energies can also be reshaped into multi-dimensional tensors that have
exactly the same dimensions as their Green's functions
counterparts. However, the self-energies cannot be computed
independently, one energy-momentum or frequency-momentum pair
depending on many others, as defined in
Eqs.~(\ref{eq:3}-\ref{eq:5}) and depicted in Fig.~\ref{fig:gf_sse}. Furthermore, ${\mathbf \Sigma}^{\gtrless
  S}(E,k_z)$ is a function of ${\mathbf D}^{\gtrless}(\omega,q_z)$,
while ${\mathbf G}^{\gtrless S}(E,k_z)$ is needed to calculate
${\mathbf \Pi}^{\gtrless}(\omega,q_z)$.

To obtain the electrical and energy currents that flow through a given
device and the corresponding charge density,
Eqs.~(\ref{eq:1}-\ref{eq:2}) (GF) and Eqs.~(\ref{eq:3}-\ref{eq:5})
(SSE) must be iteratively solved until convergence is reached, and all
GF contributions must be accumulated \cite{stieger}. The algorithm
starts by setting
${\mathbf\Sigma}^\gtrless(E,k_z)$=${\mathbf\Pi}^\gtrless(\omega,q_z)$=0
and continues by computing all GFs under this condition. The latter
then serve as inputs to the next phase, where the SSE are evaluated
for all ($k_z$,$E$) and ($q_z$,$\omega$) pairs. Subsequently, the SSE
matrices are fed into the GF calculation and the process repeats
itself until the GF variations do not exceed a pre-defined
threshold. In terms of HPC, the main challenges reside in the
distribution of these quantities over thousands of compute units, the
resulting communication-intensive gathering of all data to handle the
SSE phase, and the efficient solution of Eqs.~(\ref{eq:3}-\ref{eq:5})
on hybrid nodes, as they involve many small matrix
multiplications. Typical simulation parameters are listed in
Table~\ref{tab:common-values}.

\vspace{-2mm}
\subsection{Current State of the Art}

There exist several atomistic quantum transport simulators \cite{ferrer2014gollum,groth2014kwant,nemo5,atk,vides,niquet,omen} that can model the
characteristics of nano-devices.
Their performance is summarized in Table \ref{tab:competitors}, where their
estimated maximum number of atoms that can be simulated for a given physical
model is provided.
Only orders of magnitude are shown, as these quantities depend on the device
geometries and bandstructure method.
It should be mentioned that most tools are limited to tight-binding-like (TB)
Hamiltonians, because they are computationally cheaper than DFT ones ($N_{orb,TB} < N_{orb,DFT}$
and $N_{B,TB} \ll N_{B,DFT}$).
This explains the larger systems that can be treated with TB.
However, such approaches lack accuracy when it comes to the exploration of
material stacks, amorphous layers, metallic contacts, or interfaces.
In these cases, only DFT ensures reliable results, but at much higher computational cost.

To the best of our knowledge, the only tool that can solve
Eqs.~(\ref{eq:1}) to (\ref{eq:5}) self-consistently, in structures
composed of thousands of atoms, at the DFT level is OMEN, a two times Gordon Bell Prize finalist
\cite{sc11,sc15}.\footnote{\hl{Previous achievements: development 
of parallel algorithms to deal with ballistic transport}
(Eq.~(\ref{eq:1}) alone) \hl{expressed in a tight-binding} (SC11~\cite{sc11}) \hl{or DFT}
(SC15~\cite{sc15}) \hl{basis.}}
The code is written in C++, contains 90,000
lines of code in total, and uses MPI as its communication protocol. 
Some parts of it have
been ported to GPUs using the CUDA language and taking advantage of
libraries such as cuBLAS, cuSPARSE, and MAGMA. The electron-phonon
scattering model was first implemented based on the tight-binding
method and a three-level MPI distribution of the workload (momentum,
energy, and spatial domain decomposition). A first release of the
model with equilibrium phonon (${\mathbf \Pi}$=0) was validated up to
95k cores for a device with $N_A$=5,402, $N_B$=4, $N_{orb}$=10,
$N_{k_z}$=21, and $N_E$=1,130. These runs showed that the application
can reach a parallel efficiency of 57\%, when going from 3,276 up to
95,256 cores, with the SSE phase consuming from 25\% to 50\% of the
total simulation times. The reason for the SSE increase could be
attributed to the communication time required to gather all
Green's Function inputs for Eq.~(\ref{eq:3}), which grew from 16 to
48\% of the total simulation time~\cite{sc10} as the number of cores
went from 3,276 to 95,256. 

After extending the electron-phonon scattering model to DFT and adding
phonon transport to it, it has been observed that the time spent in
the SSE phase (communication and computation) explodes. Even for a
small structure with $N_A$=2{,}112, $N_{orb}$=4, $N_{k_z}$=$N_{q_z}$=11,
$N_E$=650, $N_{\omega}$=30, and $N_B$=13, 95\% of the total simulation
time is dedicated to SSE, regardless of the number of used
cores/nodes, among which $\sim$60\% for the communication between
the different MPI tasks. To simulate self-heating in realistic FinFETs
($N_A>$10,000), with a high accuracy ($N_{k_Z}>$20, $N_E>$1{,}000), and within
reasonable times (a couple of minutes for one GF-SSE iteration at
machine scale), the algorithms involved in the solution of
Eqs.~(\ref{eq:1}) to (\ref{eq:5}) must be drastically improved: as
compared to the state of the art, an improvement of at least one order of
magnitude is needed in terms of the number of atoms that can be
handled, and two orders of magnitude for what concerns the computational
time.

\section{Data-Centric Parallel Programming}

Communication-Avoiding (CA) algorithms~\cite{demmel13ca,writeav} are defined as algorithm variants and schedules 
(orders of operations) that minimize the total number of performed memory loads and stores,
achieving lower bounds in some cases. 
To achieve such bounds, a subset of those algorithms is \textit{matrix-free}\footnote{The term is derived from solvers that do not need to store the entire matrix in memory.}, potentially reducing communication at the expense of recomputing parts of the data on-the-fly.
A key requirement in modifying an algorithm to achieve communication avoidance is to explicitly
formulate its data movement characteristics. 
The schedule can then be changed by reorganizing the data flow to minimize the sum of
accesses in the algorithm.
Recovering a Data-Centric (DaCe) view of an algorithm, which makes movement explicit 
throughout all levels (from a single core to the entire cluster), is thus the path 
forward in scaling up the creation of CA variants to more complex algorithms and multi-level memory hierarchies as one.

\begin{figure}[t]
	\centering
	\includegraphics[width=\linewidth,page=3]{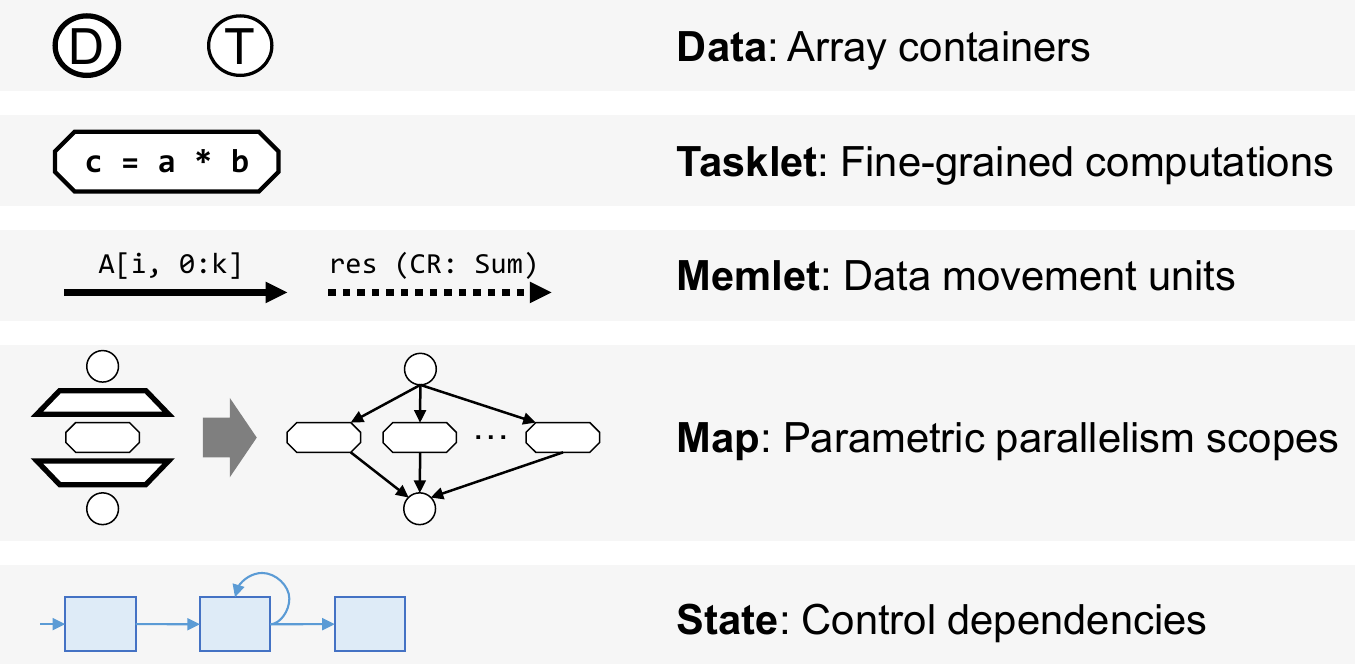}
	\vspace{-2em}
	\caption{SDFG concepts and syntax.}
	\vspace{-1em}
	\label{fig:sdfg}
\end{figure}

DaCe defines a development workflow where the original algorithm is independent from its
data movement representation, enabling symbolic analysis and transformation of the latter
without modifying the scientific code. 
This way, a CA variant can be formulated and developed by a performance engineer, while 
the original algorithm retains readability and maintainability.
At the core of the DaCe implementation is the Stateful DataFlow multiGraph (SDFG) \cite{sdfg}, an intermediate representation that encapsulates data movement and can be generated from high-level code in Python.
The syntax (node and edge types) of SDFGs is listed in Fig.~\ref{fig:sdfg}. The workflow is as follows:
The domain scientist designs an algorithm and implements it as linear algebra operations (imposing dataflow implicitly), or using Memlets and Tasklets (specifying dataflow explicitly).
This implementation is then parsed into an SDFG, where performance engineers may apply
graph transformations to improve data locality. 
After transformation, the optimized SDFG is compiled to machine code for performance 
evaluation. It may be further transformed interactively and tuned for different target platforms and memory hierarchy characteristics.

\begin{figure}[t]
	\centering
	\includegraphics[height=1.68in,page=2]{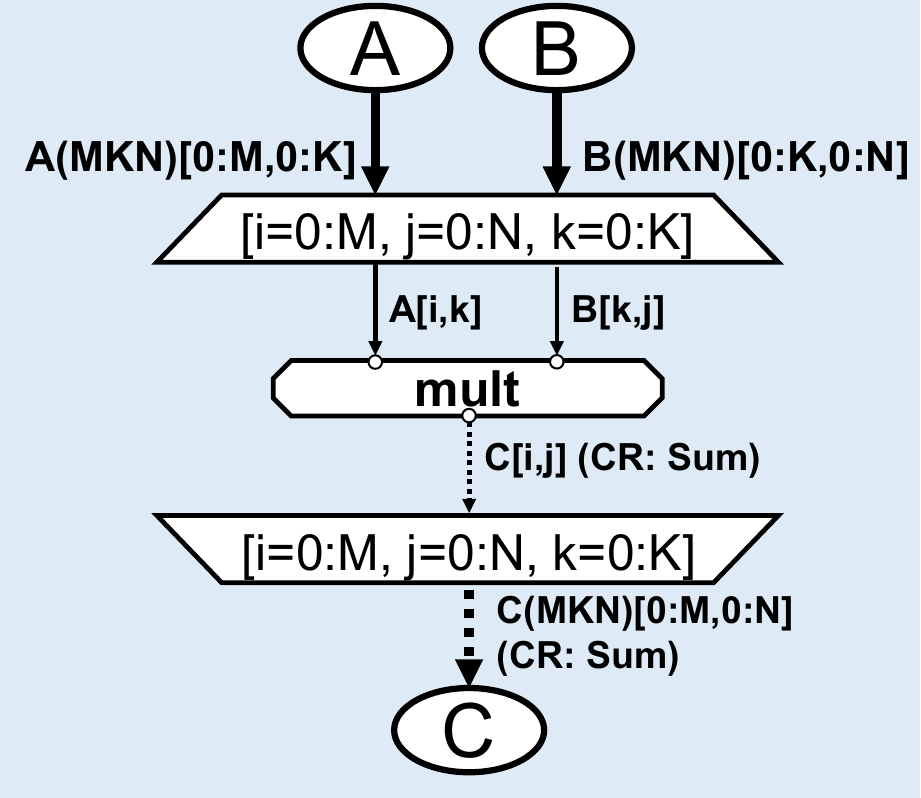}
	\vspace{-1em}
	\caption{Matrix multiplication in DaCe.}
	\vspace{-1em}
	\label{fig:dacemm}
\end{figure}

An example of a na\"{i}ve matrix multiplication SDFG (\texttt{C = A @ B} in Python) is shown in Fig.~\ref{fig:dacemm}. 
In the figure, we see that data flows from Data nodes \texttt{A} and \texttt{B} through a Map scope. 
This would theoretically expand to \texttt{M*N*K} multiplication Tasklets (\texttt{mult}), where the contribution of each Tasklet (i.e., a multiplied pair) will be summed in Data node \texttt{C} (due to conflicting writes that are resolved by \texttt{CR: Sum}). The Memlet edges define all data movement, which is seen in the input and output of each Tasklet, but also entering and leaving the Map with its overall requirements (in brackets) and number of accesses (in parentheses).
The accesses and ranges are symbolic expressions, which can be summed to obtain the algorithm's data movement characteristics.
The SDFG representation allows the performance engineer to add transient (local) arrays, reshape and nest Maps (e.g., to impose a tiled schedule), fuse multiple scopes, map computations to accelerators (GPUs and FPGAs), and other transformations that may modify the overall number of accesses.

\begin{figure}[h]
	\vspace{-1em}
	\begin{lstlisting}[numbers=left,backgroundcolor=\color{lightyellow}]
# Declaration of symbolic variables
Nkz, NE, Nqz, Nw, N3D, NA, NB, Norb = (
    dace.symbol(name)
    for name in ['Nkz', 'NE', 'Nqz', 'Nw',
                 'N3D', 'NA', 'NB', 'Norb'])

@dace.program
def sse_sigma(neigh_idx: dace.int32[NA, NB],
              dH: dace.float64[NA, NB, N3D, Norb, Norb],
              G: dace.complex128[Nkz, NE, NA, Norb, Norb],
              D: dace.complex128[Nqz, Nw, NA, NB, N3D, N3D],
              Sigma: dace.complex128[Nkz, NE, NA, Norb, Norb]):

    # Declaration of Map scope
    for k, E, q, w, i, j, a, b in dace.map[0:Nkz, 0:NE,
                                           0:Nqz, 0:Nw,
                                           0:N3D, 0:N3D,
                                           0:NA, 0:NB]:
        f = neigh_idx[a, b]
        dHG = G[k-q, E-w, f] @ dH[a, b, i]
        dHD = dH[a, b, j] * D[q, w, a, b, i, j]
        Sigma[k, E, a] += dHG @ dHD

if __name__ == '__main__':
    # Initialize symbolic variables
    Nkz.set(21)
    NE.set(1000)
    ...
    # Initialize input/output arrays
    idx = numpy.ndarray((NA.get(), NB.get()), numpy.int32)
    ...
    # Call dace program
    sse_sigma(neigh_idx=idx, dH=dH, G=G, D=D, Sigma=Sigma)\end{lstlisting}
    \vspace{-1em}
	\caption{$\Sigma^\gtrless$ computation in Python}
	\vspace{-1em}
	\label{fig:frontend}
\end{figure}

Fig.~\ref{fig:frontend} \hl{shows the computation of $\Sigma^\gtrless$ in DaCe, implemented
with linear algebra operations in a Python-based frontend, while the resulting SDFG is presented in} Fig.~\ref{fig:sse-initial}.
\hl{Symbolic variables, such as the number of atoms, momentums and energies, are
declared in lines 2-5.}
\hl{The \texttt{dace.program} decorator (line 7) is used to define the function to be converted to an SDFG. 
Type annotations in the function signature (lines 8-12) are used to define the datatype
and shape of the input and output arrays.}
\hl{For-loop statements using the \texttt{dace.map} iterator (lines 15-18) define a Map scope.
Linear algebra operations (lines 20-22) are automatically parsed to Tasklets.
The latter can be subsequently lowered to nested SDFGs that implement
these operations in fine-grained dataflow, such as the matrix multiplication SDFG
in} Fig.~\ref{fig:dacemm}. \hl{Alternatively, they can be mapped to optimized BLAS calls
when generating code.
The DaCe program is executed through Python host code (lines 24-33), where the
symbolic variables, input and output arrays are initialized.}

Our main innovation in optimizing the OMEN QT simulator lies in
the use of the DaCe parallel programming framework.
In the following section, we show how the data-centric view provided by DaCe
is used to identify and implement a \textit{tensor-free} CA variant of OMEN,
achieving optimal communication for the first time
in this scientific domain.

\section{Transforming OMEN}

To understand the dataflow of the OMEN implementation, its 90,000 lines of code, or 15,798 lines\footnote{generated using David A. Wheeler's 'SLOCCount'.} of core RGF and SSE computations can be examined. 
Alternatively, the SDFG could be used to obtain a hierarchical view of the application, where States and Map scopes can be collapsed. 
A deeper dive allows optimization of certain regions.
Below, we take a methodological top-down approach to transform the OMEN SDFG, 
starting from its high-level decomposition, which generates the communication,
through individual computational kernels, to small-scale linear algebra operations.
We instrument the code in order to find bottlenecks and critical subgraphs to ``cut out'' and transform.
Furthermore, we support our decisions with communication and performance models obtained using
the data-centric representation.

\begin{figure}[b]
	\centering 
	\includegraphics[width=\linewidth, page=2]{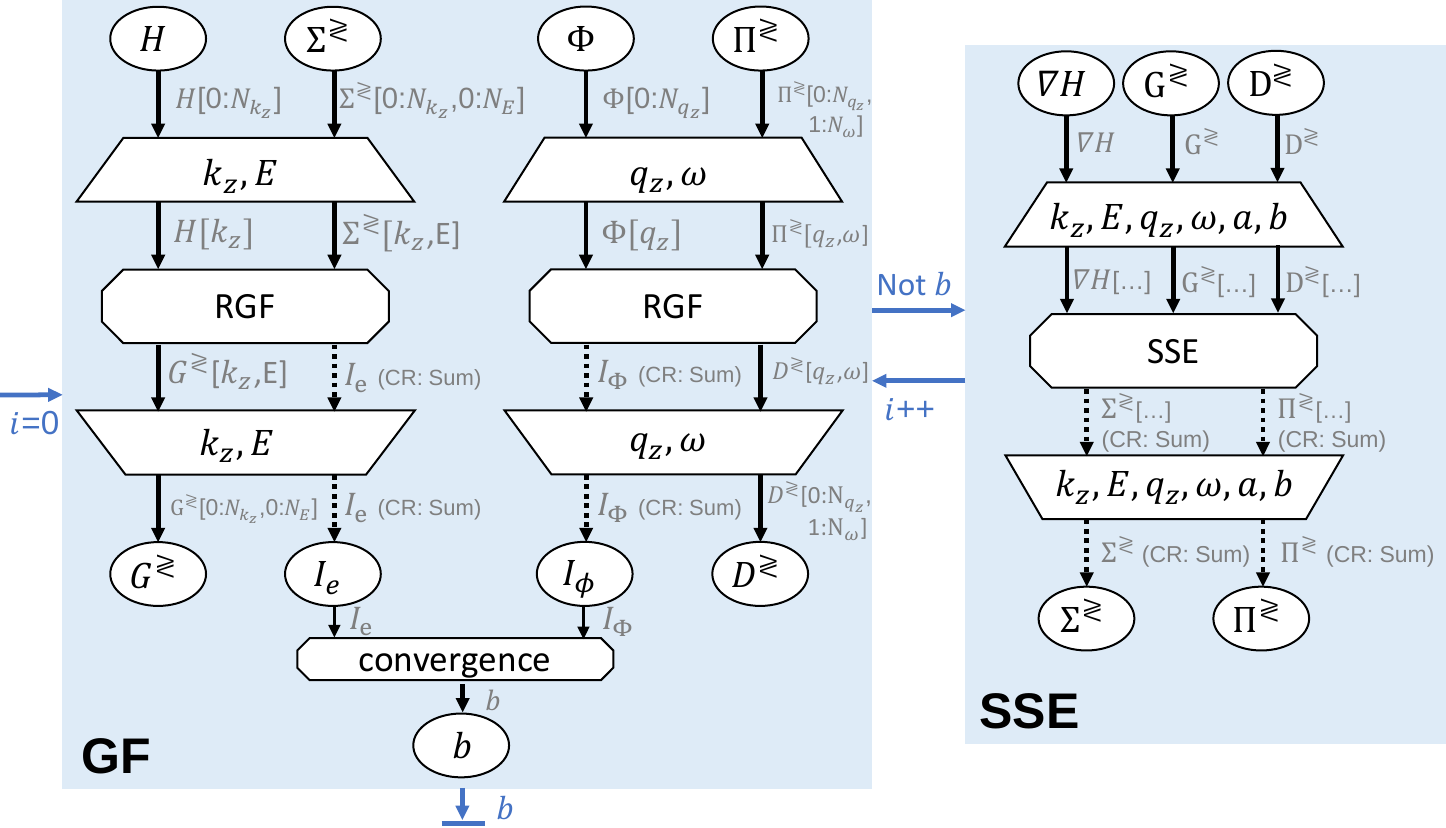}
	\vspace{-2em}
	\caption{SDFG of QT simulation: high-level performance engineer view of the problem.}
	\label{fig:dace-view}
\end{figure}

The top-level view of the QT simulation algorithm can be seen in Fig.~\ref{fig:dace-view}. 
The SDFG shows that the simulation iterates over two states, GF and SSE.
The former computes the Green's Functions, boundary conditions, and the electrical current. The state consists of two concurrent Maps, one for the electrons and one for the phonons (\S~\ref{sec:problem}).
The SSE state computes the scattering Self-Energies ${\mathbf \Sigma}^\gtrless$ and ${\mathbf \Pi}^\gtrless$.
At this point, we opt to represent the RGF solvers and SSE kernel as Tasklets, i.e., collapsing their dataflow, so as to focus on high-level aspects of the algorithm. %, because they operate on all the atoms for a specific energy-momentum pair.
This view indicates that the RGF solver cannot compute the Green's Functions for a specific atom separately from the rest of the material (operating on all atoms for a specific energy-momentum pair), and that SSE 
outputs the contribution of a specific
$\left(k_z, E, q_z, \omega, a, b\right)$ point to ${\mathbf \Sigma}^\gtrless$ and ${\mathbf \Pi}^\gtrless$.
These contributions are then accumulated to the output tensors, as indicated by the dotted Memlet edges.
The accumulation is considered associative; therefore the map can compute all dimensions of the inputs and outputs in parallel.

\subsection{Communication Avoidance}\label{sec:ca}

\begin{figure}[t]
	\centering
	\includegraphics[width=\linewidth, page=14]{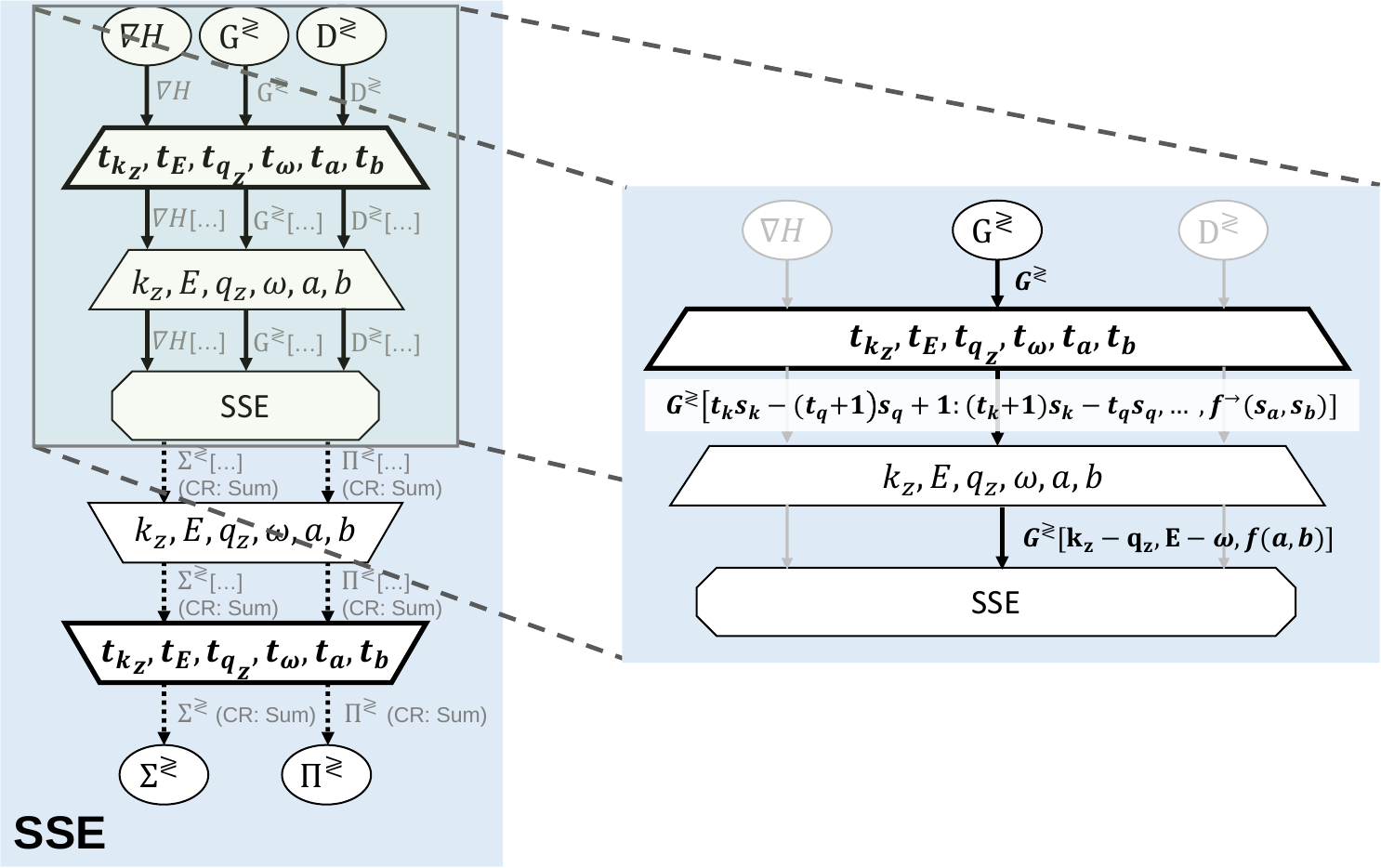}
	\vspace{-1em}
	\caption{Map-tiling SSE (left) and resulting Memlets (right).}
	\vspace{-1em}
	\label{fig:ssetiling}
	%	\end{subfigure}\qquad
	%	\begin{subfigure}[b]{.45\linewidth}
	%		\centering
	%		\includegraphics[height=1.6in, page=19]{figures/omen_figures_2.pdf}
	%		\caption{Memlet propagation for one of the ${\mathbf G}^\gtrless$ accesses.}
	%		\label{fig:sse-prop}
	%	\end{subfigure}
	%	\vspace{-1em}
	%	\caption{High-level SSE transformation.}
	%	\vspace{-1em}
	%	\label{fig:sse-hl}
\end{figure}

The applications shown in Table~\ref{tab:competitors}, including OMEN, have been developed mainly by domain scientists, and thus use the ``natural'' decomposition construction of momentum points and energies, as shown in Fig.~\ref{fig:dace-view}.
As a result, the communication scheme for SSE in OMEN is split to $N_{q_z}N_{\omega}$ rounds.
In each round:
\begin{itemize}
\item The phonon Green's Functions ${\mathbf D}^\gtrless(\omega, q_z)$ are broadcast to all processes;
\item Each process iterates over its assigned electron Green's Functions ${\mathbf G}^\gtrless(E, k_z)$,
and sends the corresponding ${\mathbf G}^\gtrless(E\pm\hbar\omega, k_z+q_z)$ to the processes that need them;
\item Each process iterates over its assigned electron Green's Functions ${\mathbf G}^\gtrless(E, k_z)$
and receives the corresponding ${\mathbf G}^\gtrless(E\pm\hbar\omega, k_z-q_z)$;
\item The partial phonon self-energies $\Pi_p^\gtrless(\omega, q_z)$ produced by each process are reduced
to ${\mathbf \Pi}^\gtrless(\omega, q_z)$. 
\end{itemize}
Based on the above, we make the following observations:
\begin{itemize}
\item The full 6-D tensors ${\mathbf D}^\gtrless$ are broadcast to all processes;
\item The full 5-D tensors ${\mathbf G}^\gtrless$ are replicated through point-to-point communication $2N_{q_z}N_{\omega}$ times.
\end{itemize}
We use DaCe to transform the SSE state and find optimal data distributions and communication schemes in the following manner:
First, we tile the SSE map (Fig.~\ref{fig:ssetiling} left, differences highlighted in bold) across all dimensions,
with the intention of assigning each tile to a different process.
The tiling graph transformation splits a map to two nested ones, where each dimension
of the original map is partitioned to $n_d$ approximately equal ranges of size $s_d$.
For example, the electron momentum dimension is partitioned to $n_{k_z}$ ranges of size $s_{k_z}$ each. The corresponding symbol $t_{k_z}$ in the outer map spans the partitions, whereas the inner symbol $k_z$ takes values in the range $\left[t_{k_z}s_{k_z}, \left(t_{k_z}+1\right)s_{k_z}\right)$.
Likewise, $q_z$ iterates over $\left[t_{q_z}s_{q_z}, \left(t_{q_z}+1\right)s_{q_z}\right)$.

Subsequently, the DaCe framework propagates the data access expressions in Memlets from the Tasklets outwards, through scopes.
DaCe automatically computes contiguous and strided ranges, but can only over-approximate some irregular accesses. In these cases, performance engineers can manually provide
the additional information to the SDFG, creating new optimization opportunities.

In particular, the propagation of the access ${\mathbf G}^\gtrless[k_z-q_z,E-\omega,f(a,b)]$
is shown in Fig.~\ref{fig:ssetiling} (right).
The propagated range of the index expression $k_z-q_z$ is computed automatically as
$[t_{k_z}s_{k_z}-\left(t_{q_z}+1\right)s_{q_z}+1,$ $\left(t_{k_z}+1\right)s_{k_z}-t_{q_z}s_{q_z})$.
The total number of accesses over this range is $s_{k_z}+s_{q_z}-1$, while the length,
which coincides with the number of unique accesses,
is $\min\left(N_{k_z}, s_{k_z}+s_{q_z}-1\right)$.
However, the expression $f(a,b)$, which represents the index of the $b$-th neighbor of atom $a$, %, which is in the range $\left[0,N_A\right)$.
is an indirection through a matrix of the atom couplings.
DaCe cannot propagate such indices and thus the performance engineer must provide a model or expression manually.

For this work, we do not to tile the dimensions of the atom neighbors.
Instead, we make use of the observation that atoms with neighboring indices are
very often neighbors in the coupling matrix.
A good approximation to the propagation of $f(a,b)$ over the range
$\left[t_as_a, \left(t_a+1\right)s_a\right)\times\left[0, N_B\right)$ is then
$\big[\min\left(0, t_as_a-\frac{N_B}{2}\right),\allowbreak\max\left(N_A, \left(t_a+1\right)s_a+\frac{N_B}{2}\right)\big)$.
The total number of accesses incr-eases to $s_aN_B$, while the length of this range is
$\min\left(N_A, s_a + N_B\right)$.

After Memlet propagation is complete, the total length of the Memlet ranges between the
two maps provides the amount of data that each process must
load/store or communicate over the network.
An \textit{\textbf{optimal communication scheme}} can subsequently be found by minimizing these expressions.
For this work, we perform exhaustive search over the feasible tile sizes.
Since the combinations of the latter are in the order of $10^6$ for most simulation parameters and number of processes,
the search completes in just a few seconds.

We demonstrate the power of the above approach by comparing the OMEN communication
scheme against partitioning the atom and electron-energy dimensions.
Using the original OMEN data distribution, each process:
\begin{itemize}
\item receives $64\frac{N_{k_z}N_E}{P}N_{q_z}N_{\omega}N_AN_{orb}^2$ bytes for the electron\\ Green's Functions ${\mathbf G}^\gtrless$;
\item sends and receives a total of $64N_{q_z}N_{\omega}N_AN_BN_{3D}^2$ bytes for the phonon Green's functions ${\mathbf D}^\gtrless$ and self-energies ${\mathbf \Pi}^\gtrless$;
\end{itemize}
where $P$ is the number of processes.
The DaCe-transformed SDFG changes the distribution of the data between the GF
and SSE states, which yields all-to-all collective operations (\texttt{alltoallv} in the MPI standard). Specifically, each process contributes:
\begin{itemize}
\item $64N_{k_z}\left(\frac{N_E}{T_E}+2N_\omega\right)\left(\frac{N_A}{T_A}+N_B\right)N_{orb}^2$ bytes for the electron Green's functions ${\mathbf G}^\gtrless$ and self-energies ${\mathbf \Sigma}^\gtrless$;
\item $64N_{q_z}N_{\omega}\left(\frac{N_A}{T_A}+N_B\right)N_BN_{3D}^2$ bytes for ${\mathbf D}^\gtrless$ and ${\mathbf \Pi}^\gtrless$.
\end{itemize}
$T_E$ and $T_A$ are the number of partitions of the energies and atoms respectively, with $P = T_ET_A$.
For ${\mathbf D}^\gtrless$ and ${\mathbf \Pi}^\gtrless$, the DaCe-based communication scheme reduces the factor
$N_AN_B$ to $\frac{N_A}{T_A}+N_B$.
In the case of ${\mathbf G}^\gtrless$, this scheme eliminates the quadratic factor over the number of momentum points exhibited by OMEN.

\subsection{Dataflow Optimizations}\label{sec:dataflow}

The data-centric view not only encompasses macro dataflow that imposes communication, but
also data movement within compute devices. We use DaCe to transform all computations in 
the communication-avoiding variant of OMEN, including the RGF algorithm, SSE, and boundary 
conditions, and automatically generate GPU code. Below we cut-out and showcase a subset of these transformations, focusing on a
bottleneck subgraph of the QT simulator, which is found within the SSE kernel: computing 
${\mathbf \Sigma}^\gtrless$ as in Eq.~(\ref{eq:3}). We note that computation of ${\mathbf \Pi}^\gtrless$ is 
transformed in a similar manner.

\begin{figure}
  \centering 
  \includegraphics[clip, height=2.4in, page=3]{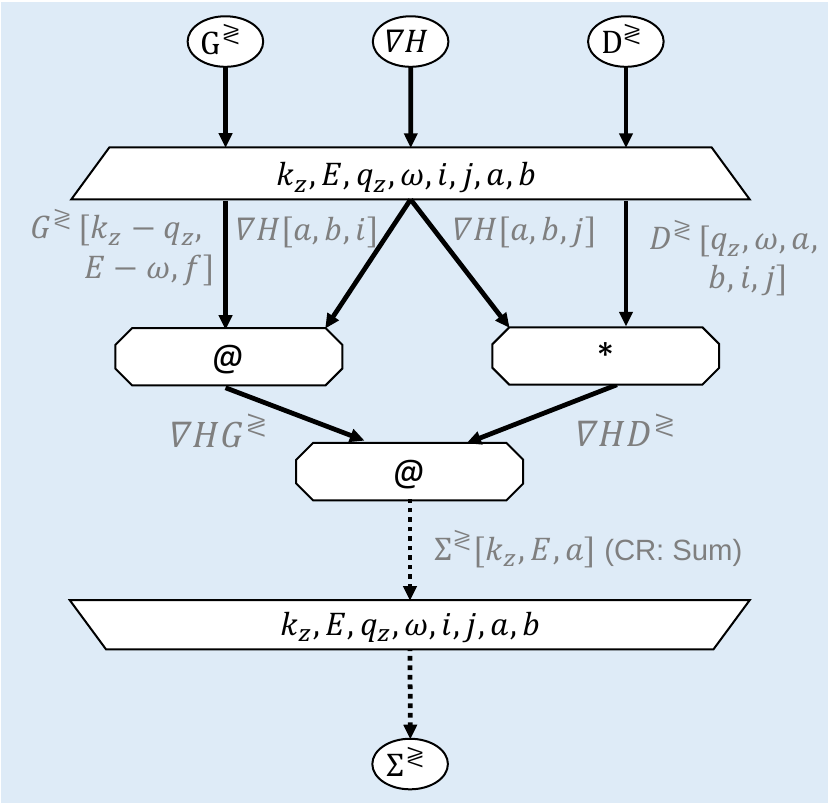}
  \vspace{-1em}
  \caption{Initial SDFG of ${\mathbf \Sigma}^\gtrless$ computation in SSE.}
  \label{fig:sse-initial}
\end{figure}

Fig.~\ref{fig:sse-initial} gives the initial representation of the computation, generated from a reference Python implementation. The inputs are:
\begin{itemize}
\item $\mathbf{G}^\gtrless$: Electron Green's Functions, a 3-D array of $N_{orb}^2$ matrices
and size $N_{k_z} \times N_E \times N_A$;
\item $\mathbf{\nabla {\mathbf H}}$: Derivative of the Hamiltonian, a 3-D array of $N_{orb}^2$ matrices
and size $N_A \times N_B \times N_{3D}$;
\item $\mathbf{D}^\gtrless$: Phonon Green's Functions, a 6-D array of scalar values
and size $N_{q_z} \times N_\omega \times N_A \times N_B \times N_{3D}^2$. Prior to the kernel, the Green's Functions have been preprocessed to contain the values
${\mathbf D}^{\gtrless ij}_{ln}(\omega,q_z)-{\mathbf D}^{\gtrless ij}_{ll}(\omega,q_z)-
{\mathbf D}^{\gtrless ij}_{nn}(\omega,q_z)+{\mathbf D}^{\gtrless ij}_{nl}(\omega,q_z)$,
as described in Eq.~(\ref{eq:3}).
\end{itemize}
The outputs are the electron self-energies ${\mathbf \Sigma}^\gtrless$, which are also a
3-D array of $N_{orb}^2$ matrices with the same dimensions as ${\mathbf G}^\gtrless$.
The SDFG consists of a map over the 8-D space
$\left[0, N_{k_z}\right)\times\left[0, N_E\right)\times\left[0, N_{q_z}\right)\times
\left[0, N_\omega\right)\times\left[0, N_{3D}\right)\times\left[0, N_{3D}\right)\times
\left[0, N_A\right)\times\left[0, N_B\right)$. For each $\left(k_z,E,q_z,\omega,i,j,a,b\right)$ point
in this space, the following computations must be performed:
\begin{enumerate}
\item The matrices at indices ${\mathbf G}^\gtrless[k_z-q_z,E-\omega,f]$ and $\nabla {\mathbf H}[a,b,i]$
are multiplied (``\texttt{@}'' symbol) and the result is stored in the temporary matrix $\nabla {\mathbf H}{\mathbf G}^\gtrless$.
The index $f$ in the array ${\mathbf G}^\gtrless$ is an indirection $f(a,b)$ in
the space $\left[0, N_A\right)$;
\item The matrix at index $\nabla {\mathbf H}[a,b,j]$ is multiplied by the scalar value
${\mathbf D}^\gtrless[q_z,\omega,a,b,i,j]$ (``$*$'' symbol) and the result is stored in the
temporary matrix $\nabla {\mathbf H}{\mathbf D}^\gtrless$;
\item The product of the temporary matrices $\nabla {\mathbf H}{\mathbf G}^\gtrless$ and $\nabla {\mathbf H}{\mathbf D}^\gtrless$
is accumulated (dotted edges) to the matrix ${\mathbf \Sigma}^\gtrless[k_z,E,a]$.
\end{enumerate}

\begin{figure}
  \centering 
  \includegraphics[clip, height=2.4in, page=4]{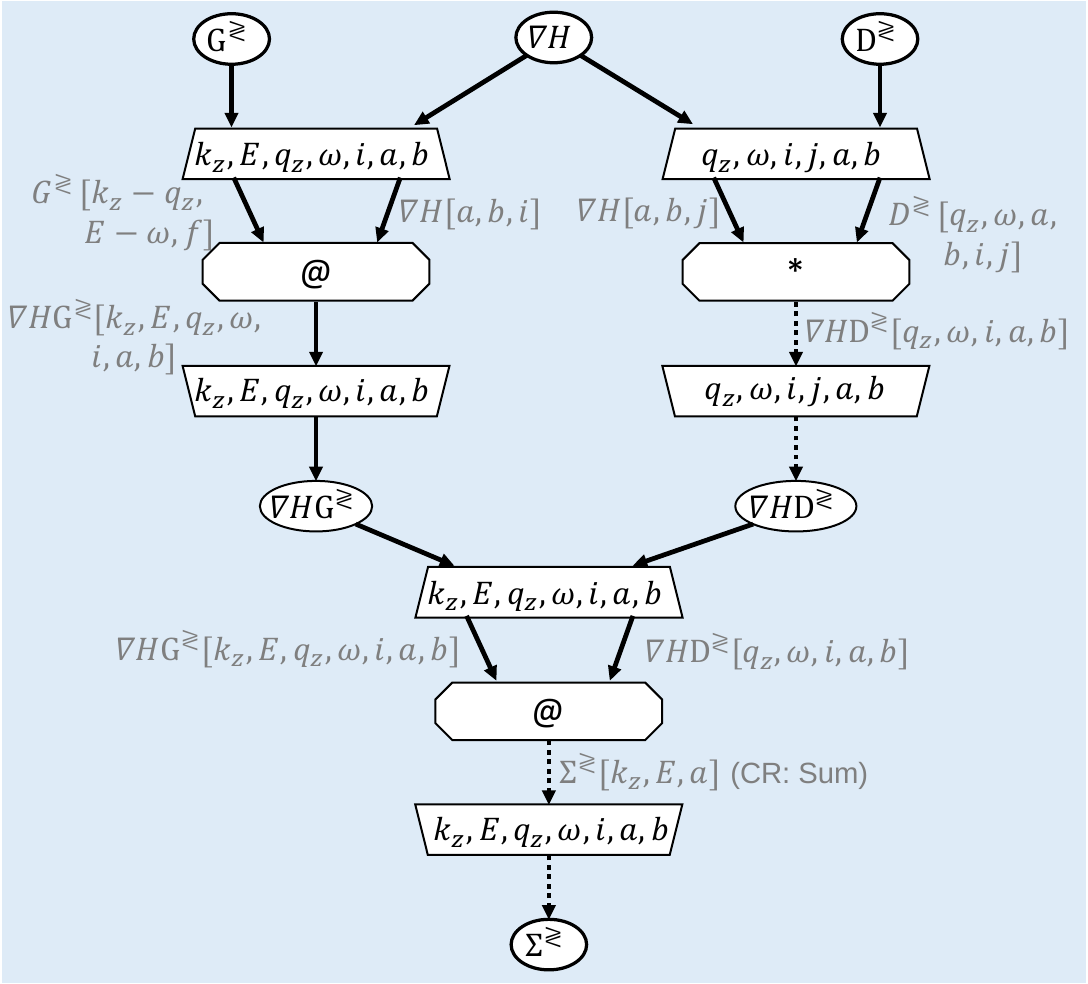}
   \vspace{-1em}
  \caption{${\mathbf \Sigma}^\gtrless$ SDFG after applying Map Fission.}
   \vspace{-1em}
  \label{fig:sse-map-fission}
\end{figure}

To optimize the SDFG, we first isolate the three computations described above.
This is achieved by applying the Map Fission (distribution) transformation, as shown in Fig.~\ref{fig:sse-map-fission}.
The transformation splits the map into three separate ones, where each one operates
over a subset of the original space.
As a result, it automatically detects that the top-left and bottom maps are independent
of the $j$ symbol, and removes it from them.
Likewise, $k_z$ and $E$ are excluded from the top-right map.
Furthermore, it substitutes the temporary matrices $\nabla {\mathbf H}{\mathbf G}^\gtrless$ and $\nabla {\mathbf H}{\mathbf D}^\gtrless$
with multi-dimensional tensors, that store all the intermediate results of two top maps.

\begin{figure*}
\centering
\begin{subfigure}[b]{.25\linewidth}
  \centering
  \includegraphics[clip, width=\linewidth, page=5]{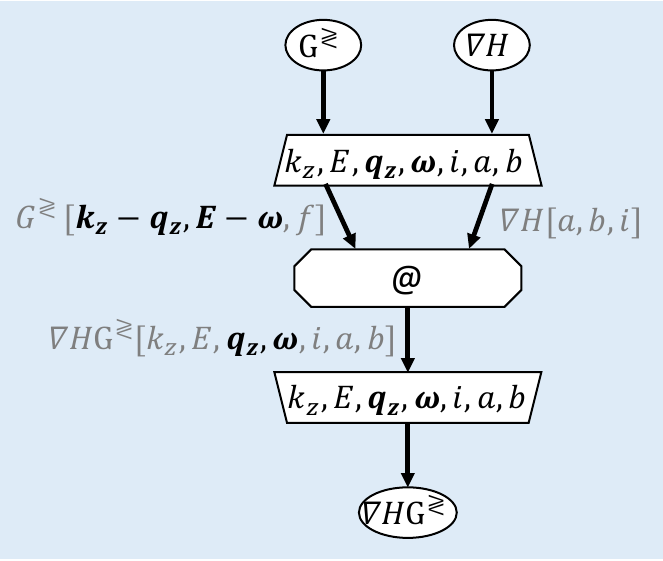}
  \caption{$\nabla {\mathbf H}{\mathbf G}^{\gtrless}$ computation.}
  \label{fig:dHGa}
\end{subfigure}\quad
\begin{subfigure}[b]{.23\linewidth}
  \centering
  \includegraphics[clip, height=1.5in, page=6]{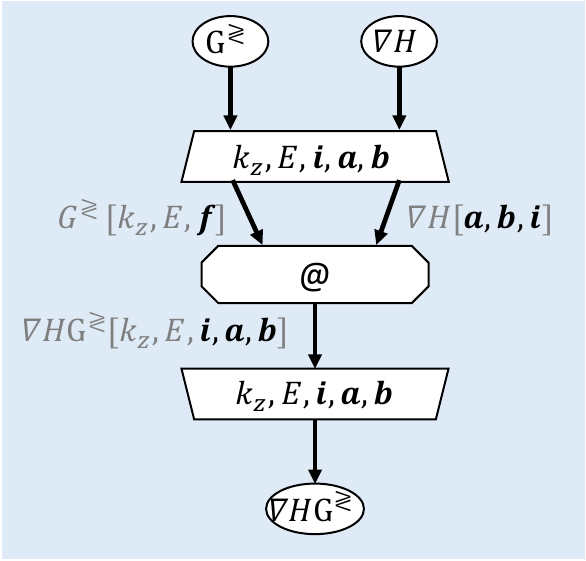}
  \caption{Redundancy removal.}
  \label{fig:dHGb}
\end{subfigure}\quad
\begin{subfigure}[b]{.23\linewidth}
  \centering
  \includegraphics[clip, height=1.5in, page=7]{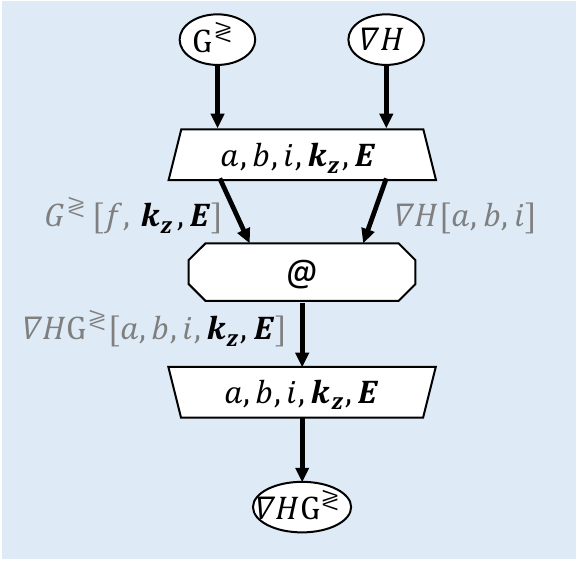}
  \caption{Data-layout transformation.}
  \label{fig:dHGc}
\end{subfigure}\quad
\begin{subfigure}[b]{.23\linewidth}
  \centering
  \includegraphics[clip, height=1.5in, page=8]{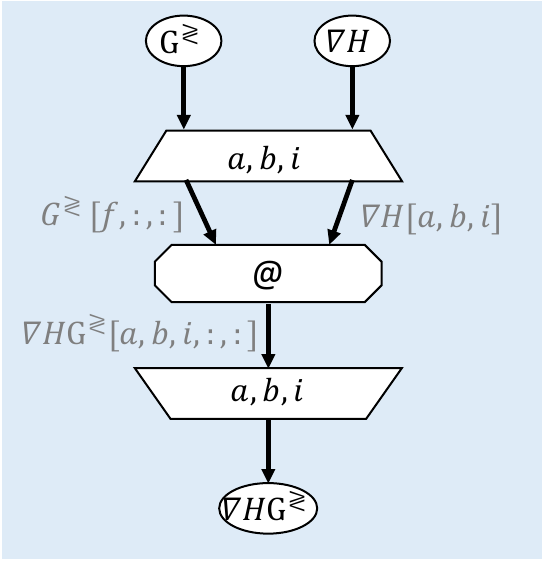}
  \caption{Multiplication fusion.}
  \label{fig:dHGd}
\end{subfigure}
\vspace{-1em}
\caption{Transformation progression on the first part of the SSE kernel (computing $\nabla {\mathbf H}{\mathbf G}^\gtrless$).}
\end{figure*}

We proceed with the optimization of the top-left map, enlarged in Fig.~\ref{fig:dHGa}.
In the subgraph, the symbols $\left(q_z, \omega\right)$ (highlighted) are only used as offsets to the 
indices $\left(k_z, E\right)$ of ${\mathbf G}^\gtrless$. Therefore, the 
subspace $\left[0, N_{k_z}\right)\times\left[0, N_E\right)$ already covers all
$\left(k_z-q_z, E-\omega\right)$ points.
The iteration over the subspace $\left[0, N_{q_z}\right)\times\left[0, N_\omega\right)$ ($q_z$ and $\omega$) results in redundant computation,
and is removed in Fig.~\ref{fig:dHGb}. The two corresponding dimensions are also removed from $\nabla {\mathbf H}{\mathbf G}^\gtrless$.

At this point, the matrices $\nabla {\mathbf H}[a,b,i]$ are used multiple times inside the map (highlighted in Fig.~\ref{fig:dHGb}),
a fact that can be exploited. However, the matrices ${\mathbf G}^\gtrless[k_z, E, f]$
are accessed irregularly, since $f$ is in this case an indirection $f(a, b)$.
This irregularity is treated by a data-layout transformation on ${\mathbf G}^\gtrless$ and
$\nabla {\mathbf H}{\mathbf G}^\gtrless$ (Fig.~\ref{fig:dHGc}). Now that the inner dimensions of
the arrays are accessed continuously over $\left(k_z, E\right)$ (highlighted), we combine
the $N_{k_z}N_E$ matrix multiplications of size $N_{orb} \times N_{orb} \times N_{orb}$ in Fig.~\ref{fig:dHGd} to a single $N_{orb} \times N_{orb} \times N_{k_z}N_EN_{orb}$ operation, with better performance characteristics.

\begin{figure*}
\begin{subfigure}[b]{.33\linewidth}
  \centering 
  \includegraphics[clip,height=1.6in, page=10]{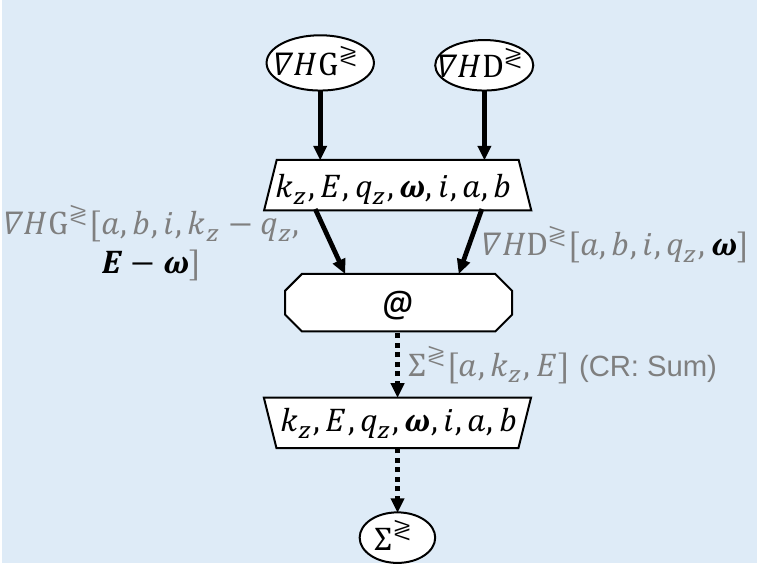}
  \caption{Continuous access over $\omega$.}
  \label{fig:sse-bottom-map}
\end{subfigure}\quad
\begin{subfigure}[b]{.3\linewidth}
  \centering 
  \includegraphics[clip,height=1.6in, page=11]{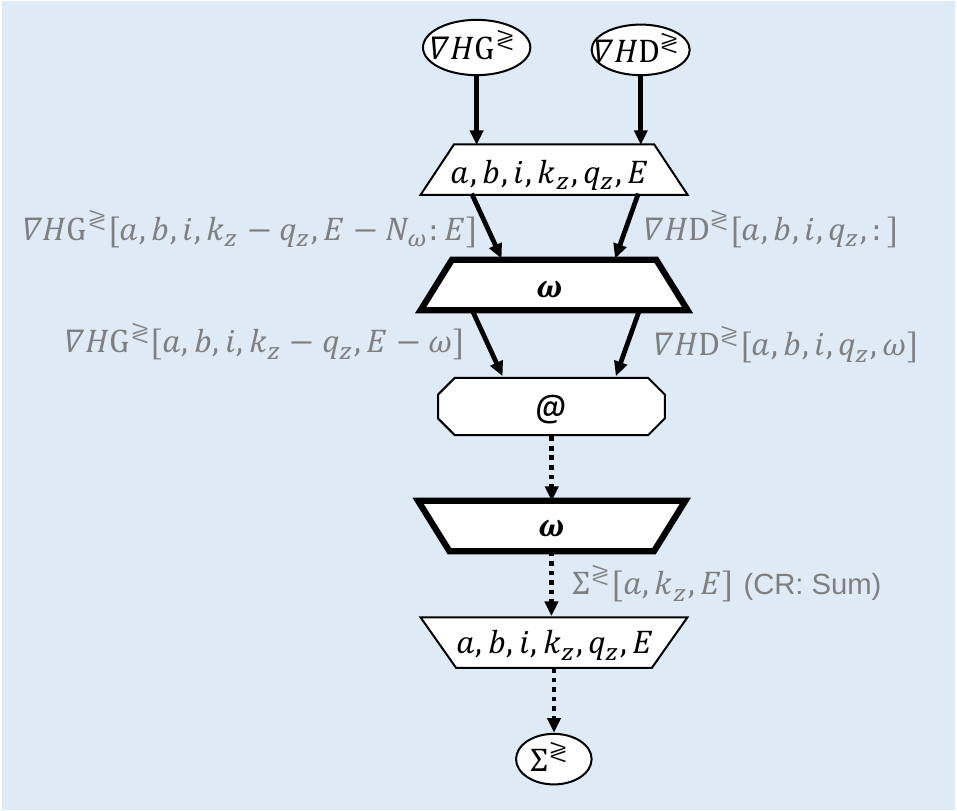}
  \caption{Map expansion.}
  \label{fig:sse-after-map-expansion}
\end{subfigure}\quad
\begin{subfigure}[b]{.33\linewidth}
  \centering 
  \includegraphics[clip,height=1.6in, page=12]{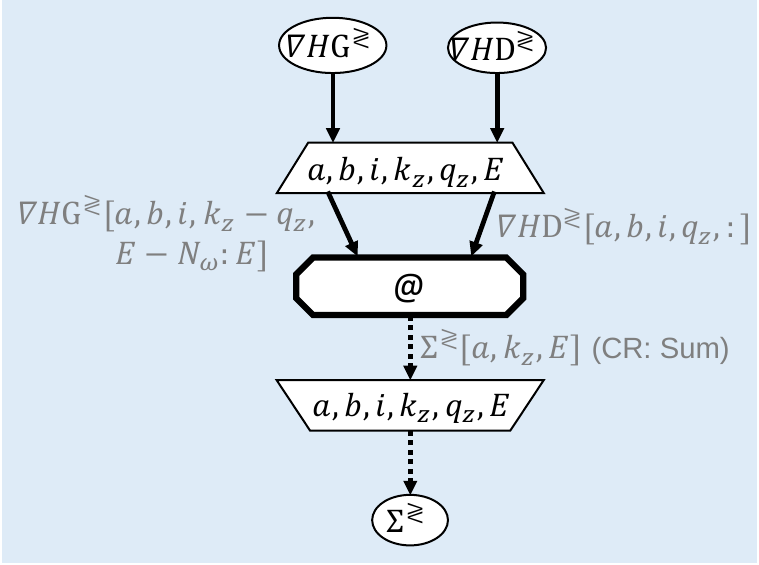}
  \caption{Nested map substitution.}
  \label{fig:sse-second-mm}
\end{subfigure}
\vspace{-1em}
\caption{Transformation progression on the third part of the SSE kernel (computing ${\mathbf \Sigma}^\gtrless$).}
\end{figure*}

Our next optimization target is the third computation (${\mathbf \Sigma}^\gtrless$) in the SSE kernel, found in the bottom map enlarged in Fig.~\ref{fig:sse-bottom-map}.
In the figure, both input tensors are accessed in a continuous manner
over $\omega$. In Fig.~\ref{fig:sse-after-map-expansion} we apply Map Expansion to create a nested map over the space $\left[0, N_\omega\right)$. 
The nested map performs the accumulation (showing only the inner indices)
${\mathbf \Sigma}^\gtrless[E]$ \texttt{+=} $\sum_\omega\{\nabla {\mathbf H}{\mathbf G}^\gtrless[E-\omega]\cdot\nabla {\mathbf H}{\mathbf D}^\gtrless[\omega]\}$,
which can be rewritten as
${\mathbf \Sigma}^\gtrless[E]$ \texttt{+=} $\nabla {\mathbf H}{\mathbf G}^\gtrless[E-\omega:E]\cdot\nabla {\mathbf H}{\mathbf D}^\gtrless[:]^T$.
In Fig.~\ref{fig:sse-second-mm} we substitute the nested map with a single
$N_{orb} \times N_{orb}N_\omega \times N_{orb}$ GEMM operation, which typically performs better than the individual small matrix multiplications.

\begin{figure}[h]
  \centering 
  \includegraphics[clip,height=2.4in, page=13]{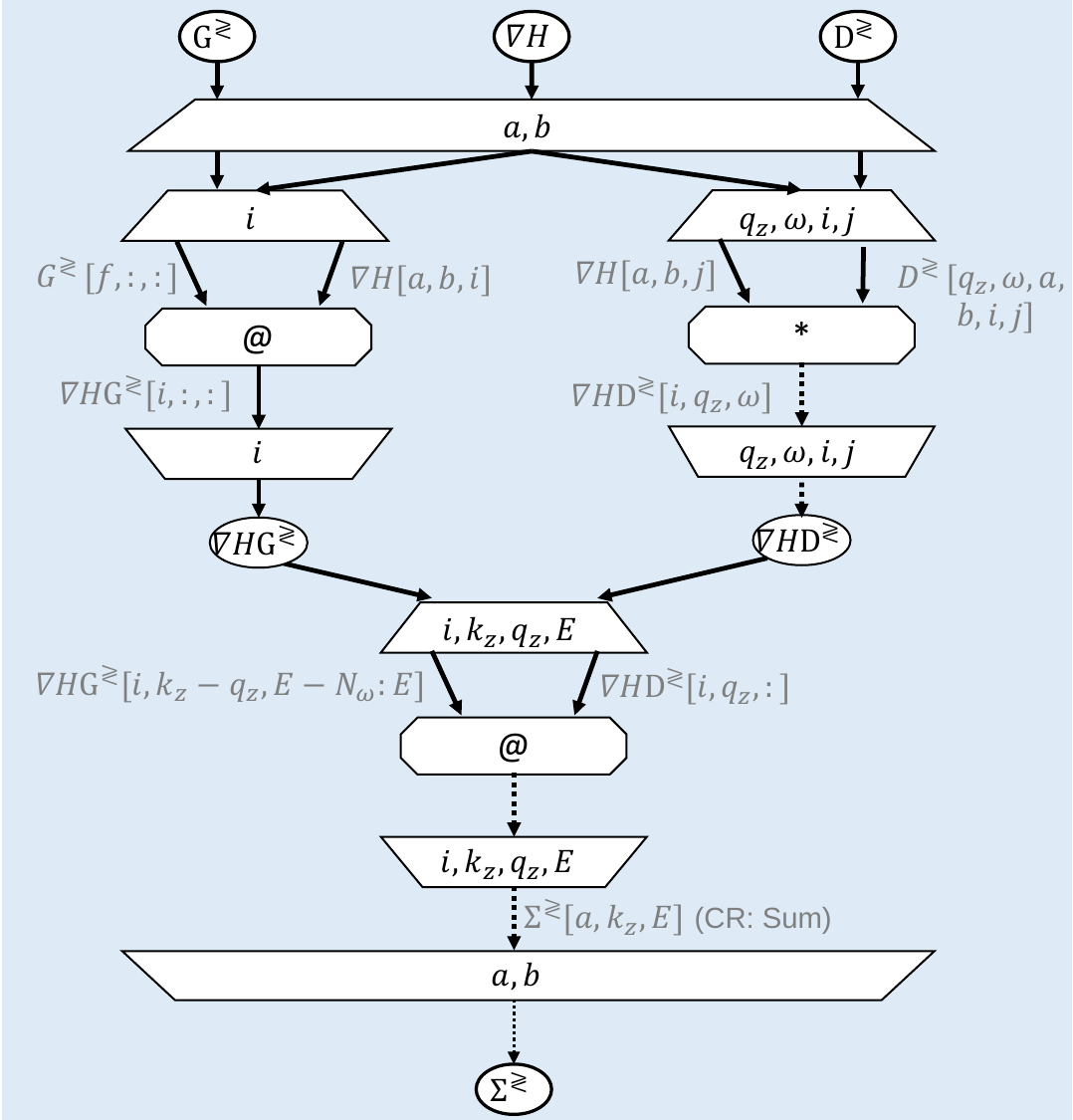}
  \vspace{-1em}
  \caption{Optimized SSE ${\mathbf \Sigma}^{\gtrless}$ SDFG.}
  \label{fig:sse-final}
\end{figure}

Since the introduced transient tensors consist of multiple dimensions, their overall size may grow rapidly.
Therefore, the last optimization step involves reducing the memory footprint of the kernel.
We achieve this in two steps. First, we expand the $\left[0, N_A\right)\times\left[0, N_B\right)$ space in each of the separate maps; and second, 
we combine the three separate (outer) maps back to a single one with Map Fusion.
The result is illustrated in Fig.~\ref{fig:sse-final}. The transformation reduces
the size of the transient arrays to only three dimensions, which are accessed for each
iteration $\left(a, b\right)$.

\subsection{Performance Model}\label{sec:perfmodel}

The majority of computations in the SDFG revolves around three kernels: (a) Contour Integration, computation of the open boundary conditions; (b) Recursive Green's Function (RGF); and (c) the SSE kernel.
The first two kernels represent most of the computational load in the GF phase,
while the SSE phase comprises the SSE kernel. 

The kernels of the GF phase involve mostly matrix multiplications. 
Therefore, the computational complexity of the RGF algorithm
is $O\left(N_A^3N_{orb}^3\right)$ for each ($E$,$k_z$)
pair and $O\left(N_{k_z}N_EN_A^3N_{orb}^3\right)$ for
the entire grid. 
Due to the GF phase kernels using both dense and sparse matrices, it is
difficult to obtain an exact flop count using analytical expressions.
We overcome this issue by counting GPU flop with the NVIDIA profiler \texttt{nvprof}, since most of the computations occur on the GPU.

For the SSE phase, described in detail in the previous section, the
complexity of the multitude of small matrix
multiplications (sized $N_{orb}\times N_{orb}$) is equal to
$O\left(N_{k_z}N_EN_{q_z}N_{\omega}N_AN_BN_{orb}^3\right)$.
Since the operations only involve dense matrices, the flop count
for the original OMEN algorithm is $64N_AN_BN_{3D}N_{k_z}N_{q_z}N_EN_{\omega}N_{orb}^3$.
The data-centric transformations performed on the algorithm reduce it to
$32N_AN_BN_{3D}N_{k_z}N_{q_z}N_EN_{\omega}N_{orb}^3 +
32N_AN_BN_{3D}N_{k_z}N_EN_{orb}^3$. Table \ref{tab:flop} shows the flop values, empirical and analytical, for a Silicon structure with $N_A=4{,}864, N_B=34, N_E=706$ and $N_{\omega}=70$, for varying $N_{k_z}$ values,
corresponding to a structure with $W=$ 2.1nm and $L=$ 35nm.

\begin{table}[t]
\vspace{0.5em}
\caption{Single Iteration Computational Load (Pflop Count)}
\vspace{-1em}
\label{tab:flop}
\begin{tabular}{ l rrrrr } 
 \toprule
 & \multicolumn{5}{c}{$\boldsymbol{N_{k_z}}$}\\\cmidrule{2-6}
 \textbf{Kernel} & 3 & 5 & 7 & 9 & 11 \\
 \midrule
Contour Integral & 8.45 & 14.12 & 19.77 & 25.42 & 31.06\\
RGF & 52.95 & 88.25 & 123.55 & 158.85 & 194.15\\
SSE (OMEN) & 24.41 & 67.80 & 132.89 & 219.67 & 328.15\\
SSE (DaCe) & 12.38 & 34.19 & 66.85 & 110.36 & 164.71\\
\bottomrule
\end{tabular}\vspace{-1em} %\\\vspace{0.5em}
\end{table}

\section{Performance Evaluation}

We proceed to evaluate the performance of the data-centric OMEN algorithm. 
Starting with microbenchmarks, we demonstrate the necessity of a high-performance implementation and that critical portions of the algorithm
deliver close-to-optimal performance on the underlying systems. 
We then measure performance aspects of OMEN and the DaCe variant on a large-scale problem consisting of 4,864 atoms, between 22 and 5,400 nodes.
Lastly, we run on the full extent of a supercomputer, measuring
the heat dissipation of a 10,240 atom nanodevice with $W=$ 4.8nm and $L=$ 35nm.
All DFT input parameters in Eqs.~(\ref{eq:1}-\ref{eq:2}) were created with CP2K
and rely therefore on Gaussian-type orbitals (GTO). A 3SP basis set was used to
model all atoms. The choice of the exchange correlation function (LDA)
has no influence on the computational efficiency.

The two systems we use are CSCS Piz Daint~\cite{daint} (6th place in June's 2019's Top500 supercomputer list) and OLCF Summit~\cite{summit} (1st place). 
Piz Daint is composed of 5,704 Cray XC50 compute nodes, each equipped with a
12-core HT-enabled (2-way SMT) Intel Xeon E5-2690 CPU with 64 GiB RAM, 
and one NVIDIA Tesla P100 GPU. The nodes communicate using Cray's Aries interconnect.
Summit comprises 4,608 nodes, each containing two IBM POWER9 CPUs (21 usable physical cores with 4-way SMT) with 512 GiB RAM and six NVIDIA Tesla V100 GPUs. The nodes are connected using Mellanox EDR 100G InfiniBand organized in a Fat Tree topology.
For Piz Daint, we run our experiments with two processes per node (sharing the GPU),
apart from a full-scale run on 5,400 nodes, where the simulation parameters do not
produce enough workload for more than one process per node.
In Summit we run with six processes per node, each consuming 7 physical cores.

We conduct every experiment at least 5 times (barring extreme-scale runs), and report the median result and 95\% Confidence Interval as error bars.

\subsection{Microbenchmarks}

Below we discuss the communication aspect of SSE, followed by computational aspects of GF.
We also evaluate the single-node performance of the different OMEN implementations.

\begin{table}
	\begin{center}
		\caption{Weak Scaling of SSE Communication Volume (TiB)}\vspace{-1em}
		\label{tab:weak-bytes}
		\small		
		\begin{tabular}{l rrrrr} 
			\toprule
			\textbf{Algorithm} & \multicolumn{5}{c}{\bf $\boldsymbol{N_{k_z}}$ (Processes)}\\\cline{2-6}
			\textbf{Variant} &3 (768)&5 (1280)&7 (1792)&9 (2304)&11 (2816)\\
			\midrule
			OMEN & 32.11 & 89.18 & 174.80 & 288.95 & 431.65\\
			DaCe & 0.54 & 1.22 & 2.17 & 3.38 & 4.86\\
			\bottomrule
		\end{tabular}\\
		$N_A = 4{,}864, N_B=34, N_{orb} = 12, N_E = 706, N_{\omega}=70$.
	\end{center}
	\vspace{-1em}
\end{table}

\subsubsection{SSE Communication Pattern}

In Tables \ref{tab:weak-bytes} and \ref{tab:strong-bytes} the total communication load for the different
implementations is shown, for a Silicon material with $N_A=4{,}864, N_B=34, N_E=706$
and $N_{\omega}=70$.
In Table \ref{tab:weak-bytes}, the number of processes increases relatively to $N_{k_z}$.
The tiling parameters (\S~\ref{sec:ca}) of the DaCe implementation are $T_E = N_{k_z}$ and $T_A=7$.
In Table \ref{tab:strong-bytes}, we fix $N_{k_z}$ to 7 and vary the number of processes.
$T_E$ is always 7 and $T_A$ is equal to $32\frac{P}{112}$. The tables both show clear advantage of using the communication-avoiding variant of the algorithm, with up to two orders of magnitude speedup.

\begin{table}[t]
	\begin{center}
		\caption{Strong Scaling of SSE Communication Volume (TiB)}\vspace{-1em}
		\label{tab:strong-bytes}
		\small		
		\begin{tabular}{l rrrrr} 
			\toprule
			\textbf{Algorithm} & \multicolumn{5}{c}{\bf Processes}\\\cline{2-6}
			\textbf{Variant} & 224 & 448 & 896 & 1792 & 2688 \\
			\midrule
			OMEN & 108.24 & 117.75 & 136.76 & 174.80 & 212.84\\
			DaCe & 0.95 & 1.13 & 1.48 & 2.17 & 2.87\\
			\bottomrule
		\end{tabular}\\
		$N_A = 4{,}864, N_B=34, N_{orb} = 12, N_{k_z}=7, N_E = 706, N_{\omega}=70$.
	\end{center}
	\vspace{-1em}
\end{table}

\begin{figure*}
	\begin{subfigure}[b]{.42\linewidth}
		\centering
		\includegraphics[width=\linewidth]{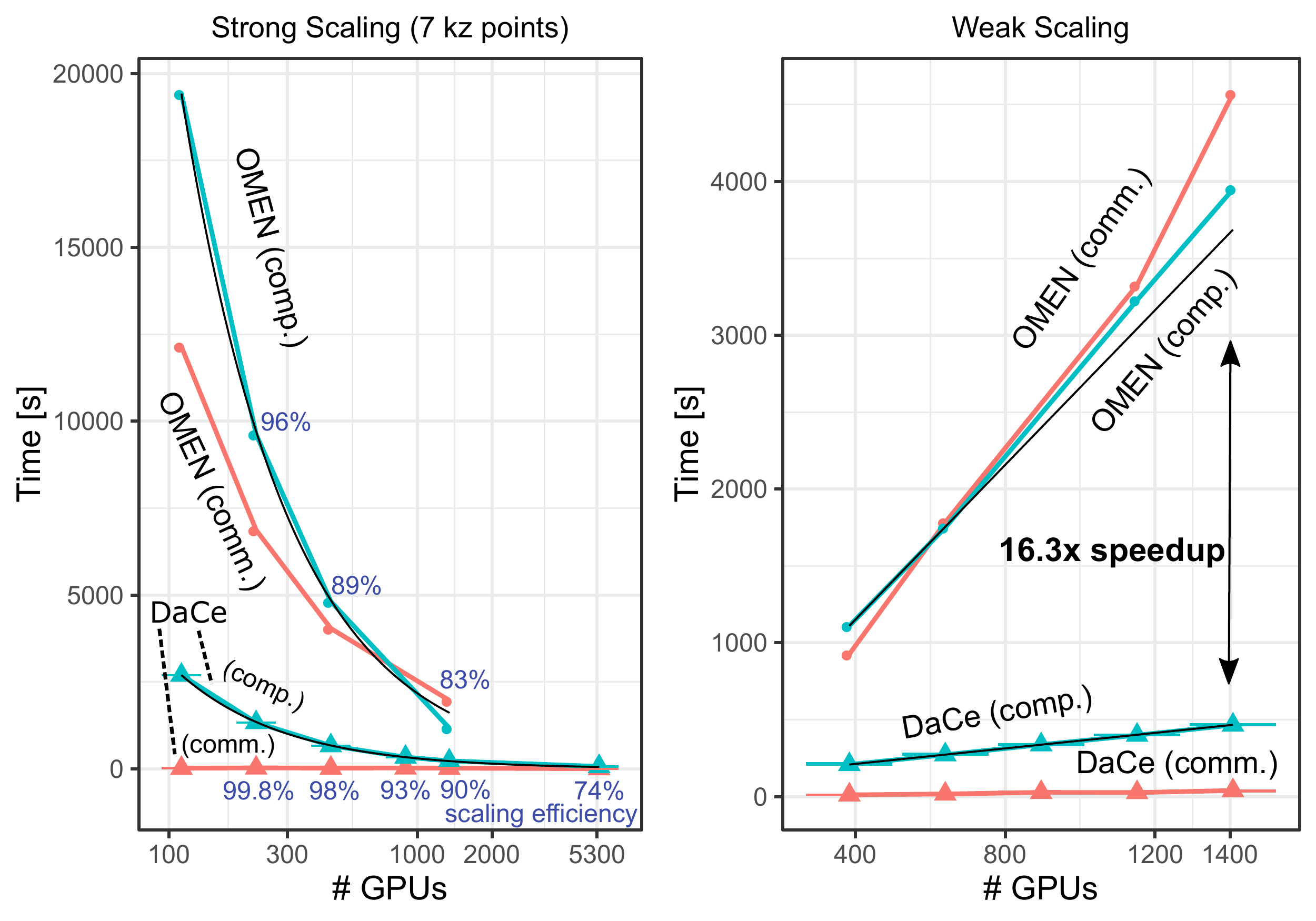}
		\vspace{-2em}
		\caption{Piz Daint}		
	\end{subfigure}\qquad\qquad
	\begin{subfigure}[b]{.42\linewidth}
		\centering
		\includegraphics[width=\linewidth]{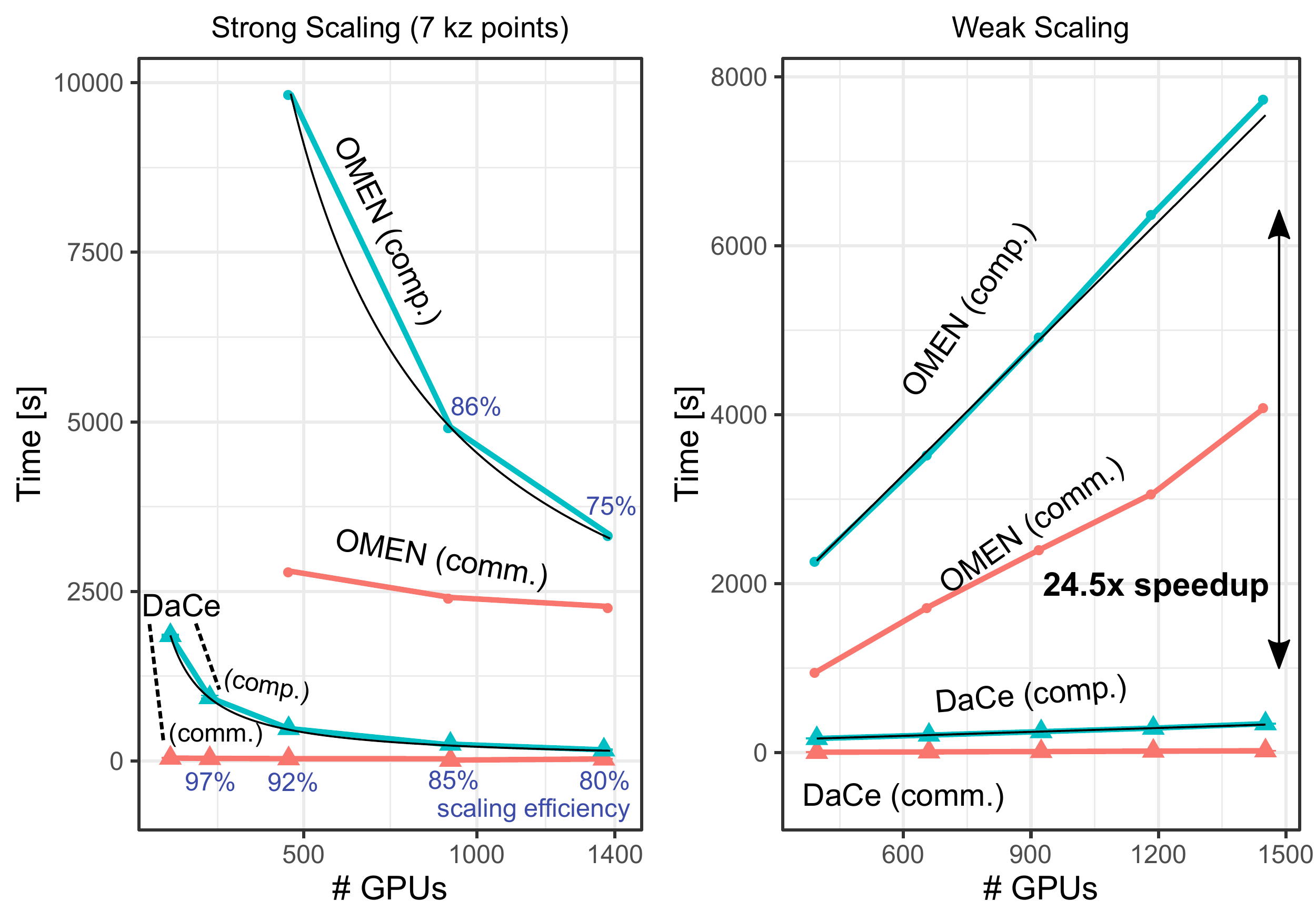}
		\vspace{-2em}
		\caption{Summit}		
	\end{subfigure}
	\vspace{-1em}
	\caption{DaCe OMEN simulation scalability ($N_a=4{,}864$, black lines:
  ideal scaling).}
	\vspace{-1em}
	\label{fig:scalability}
\end{figure*}

\subsubsection{Green's Functions and Sparsity}
Since the RGF algorithm uses a combination of sparse and dense matrices, 
there are several paths that can be taken for computing their multiplication
with each other. In particular, a common operation in
RGF is \texttt{F[n] @ gR[n + 1] @ E[n + 1]} --- multiplying two sparse blocks of the
block tri-diagonal Hamiltonian matrix (\texttt{E,F}) with a retarded Green's Functions block (\texttt{gR}). To perform this operation, one might (a) use CSR-to-dense
conversion followed by dense multiplication (\textit{Dense-MM}); (b) multiply the first CSR
matrices with the dense to obtain a dense matrix, followed by a transposed dense-CSR multiplication (\textit{CSRMM}); or (c) multiply all matrices as sparse, keeping the result (and thus \texttt{gR}) sparse (\textit{CSRGEMM}).
The first two options can be interchanged via data-centric transformations.

\begin{table}[h]
	\vspace{-0.75em}
	\caption{Sparse vs. Dense 3-Matrix Multiplication in RGF}
	\label{fig:mb:gf}
	\vspace{-1em}
	\small
	\begin{tabular}{llll}
		\toprule
		\textbf{Approach} & Dense-MM & CSRMM & CSRGEMM\\
		\midrule
		\textbf{Time [ms]} & 203.59 $\pm$ 5.95 & \textbf{47.06} $\pm$ 0.15 & 93.02 $\pm$ 0.21\\
		\bottomrule
	\end{tabular}
	\vspace{-0.75em}
\end{table}

In Table~\ref{fig:mb:gf} we study the performance of all three approaches for 
representative sizes and matrix sparsity, using cuSPARSE for the operations. All 
implementations use multiple CUDA streams (as a result of SDFG scheduling)
and thus pipeline CPU-to-GPU copies and computation. From the table, the best performance is attained with the \textit{CSRMM} approach, with 1.98--4.33$\times$ speedup. %In \textit{Dense-MM}, while the CSR-to-dense conversion is fast (taking 655 $\mu$s on average), the subsequent GEMM operation takes the lion's share of the time. This indicates that while sparse matrices

\subsubsection{Single-Node Performance}

We evaluate the performance of OMEN, the DaCe variant, and the Python reference
implementation (using the numpy module implemented over MKL), on a Silicon nanostructure with
$N_A=4{,}864, N_B=34, N_{k_z} = 3, N_E=706$ and $N_{\omega}=70$.
In Table ~\ref{tab:single-node} the runtime of the GF and SSE SDFG states is shown,
for $\frac{1}{112}$ of the total computational load, executed by a single node on Piz Daint. Although Python uses optimized routines, it exhibits very slow performance on its own. This is a direct result of using an interpreter for mathematical expressions, where
arrays are allocated at runtime and each operation incurs high overheads. This can especially be seen in SSE, which consists of many small multiplication operations. 
The table also shows that the data-centric transformations made on the Python code
using DaCe outperforms the manually-tuned C++ OMEN on both phases, where the performance-oriented reconstruction of SSE generates a speedup of 9.97$\times$.

\begin{table}[h]
	\begin{center}
		\vspace{-1em}
		\caption{Single-Node Runtime (Seconds)}\vspace{-1em}
		\small
		\label{tab:single-node}
		\begin{tabular}{l rrr} 
			\toprule
			\textbf{Phase} & \multicolumn{3}{c}{\bf Algorithm Variant}\\\cline{2-4}\addlinespace
			 & OMEN & Python & DaCe \\
			\midrule
			GF & 144.14 & 1,342.77 & \textbf{111.25} \\
			SSE & 965.45 & 30,560.13 & \textbf{96.79} \\
			\bottomrule
		\end{tabular}
	\end{center}
	\vspace{-1em}
\end{table}

\subsection{Scalability}

The communication-avoiding variant of OMEN (\textit{DaCe OMEN}) exhibits strong scaling on both supercomputers. In Fig.~\ref{fig:scalability}, we measure the runtime and scalability of a single iteration of OMEN and the DaCe variant on Piz Daint and Summit. 
For strong scaling, we set a fixed nanostructure with 4,864 atoms and $N_{k_z}=7$ (so that OMEN can treat it),
using 112--5,400 nodes on Piz Daint and 19--228 nodes (114--1,368 GPUs) on Summit.
Instead of linear scaling, we annotate ideal weak scaling (in black) with proportional
increases in the number of $k_z$ points and nodes, since
the GF and SSE phases scale differently relative to the simulation parameters
(by $N_{k_z}$ and $N_{k_z}N_{q_z}=N_{k_z}^2$ respectively).
We measure the same nanostructure with varying $k_z$ points: $N_{k_z}\in\left\{3,5,7,9,11\right\}$, using 384--1,408 nodes on Piz Daint and 
66--242 nodes (396--1,452 GPUs) on Summit.

Compared with the original OMEN, the DaCe variant is efficient, both from the
computation and communication aspects. On Piz Daint, the total runtime of the reduced-communication
variant \textbf{outperforms OMEN, the current state of the art, up to a factor of 16.3$\times$},
while the communication time \textbf{improves by up to 417.2$\times$}. On Summit,
the total runtime \textbf{improves by up to factor of 24.5$\times$}, while communication
is \textbf{sped up by up to 79.7$\times$}.

Observe that on Summit, the speedup of the computational runtime is higher than on Piz Daint.
This is the result of OMEN depending on multiple external libraries, some of which are not
necessarily optimized for every architecture (e.g., IBM POWER9). On the other hand, 
SDFGs are compiled on the target architecture and depend only on a few optimized 
libraries provided by the architecture vendor (e.g., MKL, cuBLAS, ESSL), whose implementations can be replaced by SDFGs for further tuning and transformations. 

As for scaling, on Summit DaCe OMEN achieves a total speedup of 9.68$\times$ on 12 times the nodes in the strong scaling experiment (11.23$\times$ for computation alone). Piz Daint yields similar results with 10.69$\times$ speedup. 
The algorithm weakly scales with $N_{k_z}$ on both platforms, again an order of 
magnitude faster than the state of the art. We can thus conclude that the data-centric
transformed version of OMEN is strictly desirable over the C++ version. 

\subsubsection{Extreme-Scale Run}

We run DaCe OMEN on a setup that is not possible on the original OMEN,
due to infeasible memory requirements of the algorithm. We simulate a 
10,240 atom, Silicon-based nanostructure --- \textit{a size never-before-simulated with DFT/SSE at the ab initio level} --- using the DaCe variant of OMEN.
For this purpose, we use up to 76.5\% of the Summit supercomputer: 21,150 GPUs, and run our proposed Python code with up to 21 $k_z$ points, which are necessary to produce accurate results. %which
produces accurate results and costs 7,363 petaflop/iteration.
This achieves 12\% of effective peak performance (44.5\% for the GF state and 6.2\% for the SSE one), including communication.
The simulation costs 7,363 Pflop/iteration, achieving a sustained performance of 19.71 Pflop/s (12.83\% of the effective peak), including communication.
The results are listed in Table \ref{tab:fullscale}, proving that the electro-thermal properties of nano-devices of this magnitude can be computed in under 7 minutes per iteration,
as required for practical applications. A full-scale run on Summit, with further optimizations, is described by Ziogas et al.~\cite{ziogas19}.

\begin{table}[h]
	\vspace{-0.5em}
	\caption{Summit Performance on 10,240 Atoms}
	\vspace{-1em}
	\small
	\label{tab:fullscale}
	\begin{tabular}{l r r r r r}
		\toprule
		& \multicolumn{4}{c}{\textbf{Computation}} & \textbf{Comm.}
		\\\cline{2-5}\addlinespace
		\multicolumn{1}{c}{$\boldsymbol{N_{k_z}}$} & \multicolumn{2}{c}{\textbf{GF state}} & \multicolumn{2}{c}{\textbf{SSE state}} &
		\\\cline{2-6}\addlinespace
		\textbf{(Nodes)} & \textbf{Pflop} & \textbf{Time [s]} & \textbf{Pflop} & \textbf{Time [s]}& \textbf{Time [s]}\\
        \midrule
		%\textbf{Time [s]} & 251.236 & 122.348\\
		%\midrule
		\addlinespace
		%\textbf{PFLOPS} & \multicolumn{2}{c}{\textbf{51.213}, 33.34\% of effective peak}\\
		11 (1852) & 2,922 & 75.84 & 490 & 95.46 & 44.02\\
		15 (2580) & 3,985 & 75.90 & 910 & 116.67 & 43.93\\
		21 (1763) & 5,579 & 150.38 & 1,784 & 346.56 &  121.91\\
		21 (3525) & 5,579 & 76.09 & 1,784 & 175.15 & 122.35\\
		\bottomrule
	\end{tabular}
	$N_A = 10{,}240, N_B=34, N_{orb} = 12, N_E = 1{,}000, N_{\omega}=70$.
	\vspace{-0.75em}
\end{table}

\section{Conclusions}

This paper shows that modifications to data movement alone can transform a dissipative quantum 
transport simulation algorithm to become communication-efficient. Through rigorous modeling 
made possible by a data-centric intermediate representation, and graph transformations on 
the underlying macro and micro dataflow, this work is the first to introduce communication-avoiding 
principles to a full application. The algorithm is run on two of the fastest supercomputers, where the 
performance is increased by up to \textit{two orders of 
magnitude} over the previous state of the art, measuring heat dissipation of nanodevices 
with scattering self-energies, 10,240 atoms, and 21 $k_z$ points \textit{for the first 
time}. These results were obtained from a Python source code containing 3,155 lines of 
code\footnote{Generated using David A. Wheeler's 'SLOCCount'.} and an SDFG with 2,015 
nodes, all without modifying the original operations. Applying the contributions of this 
paper on the state of the art C++ code, on the other hand, would require its complete 
rewrite, due to its tightly-coupled computation and communication modules. 

\yinyang~The presented results imply that optimizing data movement separately from the 
source code can be used to \yang~further adapt this algorithm, as well as other physics simulations, to future supercomputers;
and to \yin~~augment quantum transport simulations with additional features, without undoing 
existing optimizations, enabling better cooling system designs in future microprocessors.

\begin{acks}
This work was supported by the European Research Council
(ERC) under the European Union's Horizon 2020 programme
(grant agreement DAPP, No. 678880), by the MARVEL NCCR of the
Swiss National Science Foundation (SNSF), by SNSF grant 175479
(ABIME), and by a grant from the Swiss National Supercomputing Centre, Project No.~s876. This work used resources of the Oak Ridge
Leadership Computing Facility, which is a DOE Office of Science User
Facility supported under Contract DE-AC05-00OR22725. The authors would
like to thank Maria Grazia Giuffreda, Nick Cardo (CSCS), Don Maxwell, Christopher Zimmer, and especially Jack Wells (ORNL) for access and support of the computational resources.
\end{acks}

\vspace{-0.5em}

\bibliographystyle{ACM-Reference-Format}
\bibliography{references}
    
\end{document}